\documentclass[twocolumns]{rsos}

\usepackage{float}
\usepackage{amsmath}
\usepackage{bm}
\usepackage{color}  
\usepackage{stackrel}
\usepackage{amssymb}
\usepackage{multirow}
\usepackage{units}
\usepackage{soul}
\usepackage[table]{xcolor}
\usepackage[final]{pdfpages}

\definecolor{orange}{RGB}{253,245,230}
\newcommand{\new}[1]{{\color{black}#1}}

\begin{document}

\title{Joint assessment of density correlations and fluctuations for analysing spatial tree patterns}

\author{
Villegas, P.$^{1}$, Cavagna, A.$^{1,2}$, Cencini, M.$^{1}$, Fort, H.$^{3}$ and Grigera, T.S.$^{1,4,5,6}$}

\address{
$^{1}$Istituto dei Sistemi Complessi, Consiglio Nazionale delle Ricerche, via dei Taurini 19, 00185 Rome, Italy

$^{2}$Dipartimento di Fisica, Università Sapienza, 00185 Rome, Italy

$^{3}$Institute of Physics, Faculty of Science, Universidad de la República, Iguá 4225, Montevideo 11400 Uruguay
  
$^{4}$Instituto de Física de Líquidos y Sistemas Biológicos ---
CONICET and Universidad Nacional de La Plata, La Plata, Argentina

$^{5}$CCT CONICET La Plata, Consejo Nacional de Investigaciones Científicas y Técnicas, Argentina

$^{6}$Departamento de Física, Facultad de Ciencias Exactas, Universidad Nacional de La Plata, Argentina

}



\corres{Pablo Villegas\\
\email{pvillegas@ugr.es}}

\begin{abstract}
Inferring the processes underlying the emergence of observed patterns
is a key challenge in theoretical ecology. Much effort has
been made in the past decades to collect extensive and detailed
information about the spatial distribution of tropical rainforests, as
demonstrated, e.g., in the 50$\,$ha tropical forest plot on
Barro Colorado Island, Panama. These kind of plots have been crucial
to shed light on diverse qualitative features, emerging both at the
single-species or the community level, like the spatial aggregation or
clustering at short scales. Here, we build on the progress made in the
study of the density correlation functions applied to biological
systems, focusing on the importance of accurately defining the borders
of the set of trees, and removing the induced biases. We also pinpoint
the importance of combining the study of correlations with the scale
dependence of fluctuations in density, which are linked to
the well known empirical Taylor's power law. Density correlations and
fluctuations, in conjunction, provide an unique opportunity to
interpret the behaviors and possibly to allow comparisons between data and models. We also study such quantities in models of spatial patterns and, in
particular, we find that a spatially explicit neutral
model generates patterns with many qualitative features in common with
the empirical ones.
\end{abstract}


%


\maketitle

\section{Introduction}

Ecosystems are shaped by processes -- ecological forces, e.g., seed
dispersal, interactions among species of the same or different trophic
level, and abiotic factors, e.g., the climate, fires etc. -- occurring
on different space time scales and different level of organizational
complexity \cite{levin1992}. Typically, we can have access to the
processes only through the patterns they generate
\cite{rosenzweig1995species}. Thus, a key challenge of theoretical
ecology is to infer the underlying processes from the observed patterns
\cite{levin1992}.

Paradigmatic examples of emerging patterns in ecology are tropical
rainforests.  In such biodiversity hotspots, thousands of \new{plants}
belonging to hundreds of species coexist in relatively small areas
\cite{bermingham2005tropical}, generating complex spatial patterns of
\new{vegetation}. Such patterns can be studied at the macro-level: looking
at how the number of species or the species abundance distribution
change with the sampled area \cite{rosenzweig1995species}. Or, on a
more detailed level by studying the spatial distributions of trees
\cite{perry2006comparison,velazquez2016,MWBook}.

Ideas from statistical physics, which studies the emergence of
macroscopic properties of a system from its microscopic rules, proved
to be very fruitful to understand biological systems of high level
organizational complexity. An overarching concept to understand
the emergent properties of such systems is that of
correlation. The study of correlations has been key to understand,
e.g., the rules at the basis of collective motions in
bird flocks \cite{ACRev}, the neurons of vertebrate retina
\cite{schneidman2006weak}, or the spatial yield response in a pistachio
orchard \cite{noble2018}.

In this work we are interested in the spatial density correlations of
tree patterns, using the so-called \new{pair correlation (or radial
  distribution)} function, $g(r)$, \new{which quantifies the average
  density of trees at distance $r$ from any individual tree,
  normalized by the expected value based on the mean density of vegetation \cite{Hansen1990, MWBook}}.  In the ecological literature the
same quantity is \new{also} known as neighbourhood density function
\cite{condit2000,perry2006comparison,wiegand2004rings} and \new{it has often
used to detect} e.g. clumping of trees \cite{condit2000} which reflects
in $g(r)>1$ values.   Here,
we focus on two (mainly methodological) aspects.

The first concerns how to properly take into account the biases
induced by the borders of the set of points\cite{Stoyan1994}.  This issue (often overlooked) involves two distinct problems: knowing
the borders how to reduce the biases and, more subtle, how to properly
identify the (not necessarily convex) borders of a set of points.
Both issues can be approached with different methods
\cite{hanisch1984some,Edels1983,Edels1994,Stoyan1994,wiegand2004rings,perry2006comparison}. Motivated
by their success in coping with bird flocks \cite{AShapesAC}, we use
the Hanisch method \cite{hanisch1984some} to cure the biases and the
$\alpha-$shapes method \cite{Edels1983,Edels1994} to identify the
borders. The latter consists of a geometric algorithm to carve
out concavities from a set of points using discs of a predefined
radius, and does not appear to be widely known in the ecological
literature, with a few exceptions \cite{capinha2014predicting}.

The second aspect is that it can be difficult to interpret the results
of studying only behaviour of the (properly computed) \new{pair
  correlation} function (\new{PCF}) in isolation of other relevant
quantities. This was emphasized in a recent survey of the ecological
literature \cite{velazquez2016}, which found that, in the face of a
growing number of works on spatial point pattern analysis, and of
methodological reviews on the subject
\cite{dale2000,MWBook,wiegand2004rings, perry2006comparison,
  wiegand2009}, a large percentage of the examined studies focused
only on a single observable --- mainly the \new{PCF} (sometimes
neglecting border issues), or a related function. Here, we propose to
study, in combination with the \new{PCF}, also the way \new{spatial}
density fluctuations decay with the observation scale, as it provides
useful information, especially on the large scales. \new{In this work
  we show that such decay} is simply related to one of the most well
known (empirical) laws in ecology, namely the Taylor's power law
\cite{taylor1961} that, as far as we know, was not put in combination
with the density correlation before. This law states that the standard
deviation (of time or space fluctuations) of the population size
scales as a power $\gamma$ of the mean population. As reviewed in
\cite{Eisler2008}, such relation between fluctuations and mean is
found in a wide range of disciplines with $\gamma$ typically in the
interval $[1/2, 1]$. The value $\gamma=1/2$ characterizes the behavior
of a homogeneous random processes and of cases in which the central
limit theorem applies.  Larger values, $\gamma>1/2$, typically signal
the presence of non-trivial correlations or the effect of
heterogeneities \cite{Eisler2008}. \new{In particular, anomalous
  values of $\gamma$ have been reported for almost all the species in
  Barro Colorado in Ref.~\cite{seri2015spatial}, where a closely related
  quantity -- namely the Fano factor -- was investigated}. \new{Here
  in addition to showing the anomaly of the exponent $\gamma$, as
  border bias can alter its value, we also investigate the use of
  $\alpha$-shapes method to mitigate such bias.}

To illustrate the importance of accounting for the borders and how the
combination of density correlations and \new{spatial} fluctuations can
aid in the process of interpretation and, possibly, of model
selection, we study \new{the emergent} spatial patterns of the Barro
Colorado Island (BCI) 50-ha ($1000\times 500\,$m$^2$) plot. Such
database comprises 8 censuses (every 5 years from 1980s) of more than
$4\cdot 10^5$ trees \new{and shrubs} with diameter at breast height
larger than \new{$0.01\,$m}, belonging to about 300 species, providing
position and species for each \new{plant} \cite{condit2014}. Data and
related information can be found in \cite{bcidata}.

Aiming at a qualitative comparison with BCI data and to further
exemplify the ideas here developed, we also study density correlation
and \new{spatial} fluctuations in three reference models. The first one is a simple
heterogeneous Poisson process, which is expedient to illustrate how
inhomogeneities can give rise to large density fluctuations and
misleading behaviors of the density correlation. Secondly, we study
the Thomas Process \cite{thomas1949}, one of the simplest instances of
Poisson cluster processes \cite{Stoyan1994,MWBook}, which incorporates
the idea of offsprings dispersed by parent trees. This model and its
variants have shown particular successful fitting data
\cite{wiegand2007}. \new{But, the underlying statistical properties are a prerequisite to generate the patterns, thus putting special emphasis on the inference strategy to determine the model parameters.} However, extrapolating ecological processes from these procedures is a delicate issue as combination of different mechanisms can produce similar
patterns, as highlighted in Ref.~\cite{detto2013fitting}.

Finally we consider patterns generated by spatial individual-based
model for a community of coexisting species, where the statistical
properties emerge from the incorporated processes. There are two
alternatives for such class of models, reflecting two views on how
biodiversity is maintained. On one side, niche theory
\cite{chase2003ecological} holds species differences (in resource
exploitation, reproduction strategies, etc) responsible for their
coexistence.  On the opposite side, the neutral theory
\cite{hubbell2001,azaele2016statistical} assumes species of the same
trophic level as equivalent and sees biodiversity as a nonequilibrium
stationary state realized thanks to species influx (speciation) and
random drifting toward extinction (outflux) by competition and
demographic stochasticity.  Remarkably both theories describe well some
macro-ecological patterns of biodiversity 
\cite{chave2002comparing,tilman2004niche,azaele2016statistical} and
some consensus is emerging that both mechanisms are at play
\cite{gravel2006reconciling}.  Spatially explicit models based on niche
theory are typically defined through many parameters
\cite{tilman2004niche}, conversely, owing to species equivalence,
neutral ones need very few \cite{Durrett1996,pigolotti2018}. Without
any claim of neutrality for real data, we opted for the latter just
for the sake of simplicity.   \new{Surprisingly
this simple model gives rise to a variety of behaviors, qualitatively
similar to those observed in real data, demonstrating the richness
induced by simple mechanisms even in a neutral context.}

\new{The BCI plot has been much studied in the past. Some studies
  suggested multifractal properties of the low-canopy gaps
  \cite{manrubia1996}, simulated through cellular automata models
  \cite{sole1995_PRE} and also compatible with neutrality conditions
  equipped with long-range dispersal \cite{borda2007}. Relying on the
  study of the $g(r)$, double-cluster process have been proposed to
  explain the spatial distribution of some selected species
  \cite{wiegand2009}.  The PCF and other statistical quantities (such
  as the nearest neighbour distribution) have been also studied in
  \cite{Seri2015glocal}.  Finally, a few studies have pinpointed the
  relevance of neutral competition to generate non trivial spatial
  patterns at the single-species level
  \cite{May2015,Seri2012,Grilli2012}. However, to the best of our
  knowledge systematic studies of the possible effect of borders on
  the PCF were not thoroughly \new{conducted} previously, nor in combination with
  the scale dependence of  spatial density fluctuations (which amounts to study the Taylor law as discussed in the following).}

The material is organized as follows.  In Sec.~\ref{sec:corr} we
discuss the computation of \new{PCF} focusing on some species in BCI
plot and on the role of boundaries. \new{Spatial} density fluctuations
and Taylor's law are discussed in Sec.~\ref{sec:fluct}. Correlations
and fluctuations in the above mentioned spatial models are discussed
and qualitatively compared with data in
Sec.~\ref{sec:models}. Section~\ref{sec:end} is devoted to
conclusions. Technical material and some supporting results can be
found in Supplemental Information (SI).

\section{Density correlation for the community and for single species\label{sec:corr}}

\subsection{The \new{pair correlation} function}
\label{sec:radi-distr-funct}

We start defining our main observable, namely the \new{pair correlation (or radial distribution)}
function, $g(r)$, which is here used to probe density
correlations in the spatial distribution of trees in the BCI plot. \new{The $g(r)$ is proportional to the probability to have a tree at
distance $r$ from any given tree and describes how trees are
distributed in space at varying the lenghtscale.}
\new{In particular,} given $N$ points (trees, particles etc.) in an area $A$, the function $g(r)$ is
the number of points $dN(r)$ in the annular area between $r$ and
$r+dr$ centered on one point, averaged over all points and normalized
with the expected number of neighbours for a completely random (i.e. a
homogeneous Poisson) distribution with mean density $\rho_0=N/A$. In
formulae,  the \new{PCF} reads
\begin{equation}
g\left(r\right)=\frac{dN(r)}{2\pi \rho_0 rdr}=\frac{1}{N\rho_{0}2\pi r}\stackrel[i,j]{N}{\sum}\delta\left(r-r_{ij}\right)\,,\label{eq:CorrFunc}
\end{equation}
where $i$ and $j$ denote two points in the set of interest. \new{For a Poissonian random distribution of points one has $g(r)=1$, by definition. Conversely,} values above 1 denote clumping,
i.e. tree clustering, as generically found at short distances in
rainforests \cite{condit2000}, while values below $1$ signal the
presence of ``repulsion'' or anticorrelations. In general, $g(r)$ can
have a non-trivial spatial dependence, e.g., it has spikes on regular
lattices/crystal or broad peaks in a liquid \cite{Hansen1990},
providing information on possible processes acting at different
scales.

When computing the $g(r)$ to avoid spurious results one has to
properly take into account the presence of borders. Indeed, points
close to the borders, having less neighbours than those in bulk, can
bias the statistics.  To mitigate or remove such biases different
methods can be used. For instance, one can give different weights to
bulk or border points, limit the analysis to points belonging to a
subregion in the bulk or impose periodic boundary conditions by
replicating the study area \cite{perry2006comparison}. Here, we employ
an unbiased estimator of the number of neighbours \cite{Stoyan1994}
originally proposed by Hanisch \cite{hanisch1984some}.  For each point
$i$ we count any other points $j$ as neighbour only if its distance
from $i$, $r_{ij}$, is less than that of $i$ from
its closest border, $d_i$, thus constraining the sum over trees in
Eq.~(\ref{eq:CorrFunc}).  When using binning to group distances
together, the point $j$ is excluded if it falls on a bin that is not
completely contained within the borders.

In the following we discuss the \new{PCF} in the BCI plot both at the
community level (i.e.\ considering all trees together regardless their
species) and, for some selected species, at the single species level.
We always use the Hanisch method to avoid border bias.  However, this
requires knowledge of the borders, a non trivial problem with real
data \cite{AShapesAC}. Thus we compare two different definitions of
the borders: the edges of the rectangular plot (which for most
species approximate to the convex-hull of the set), and
the borders obtained with the $\alpha$-shapes method
\cite{Edels1983,Edels1994}.  As discussed below and detailed in the SI
(Sec.~SI-1), the advantage of $\alpha$-shapes is to provide a
geometrical criterion to remove concavities (for other methods see
\cite{wiegand2004rings}).

\subsection{Rectangular borders}
\label{sec:rectangular-borders}

\begin{figure}[t!]
\centering \includegraphics[width=1.04\columnwidth]{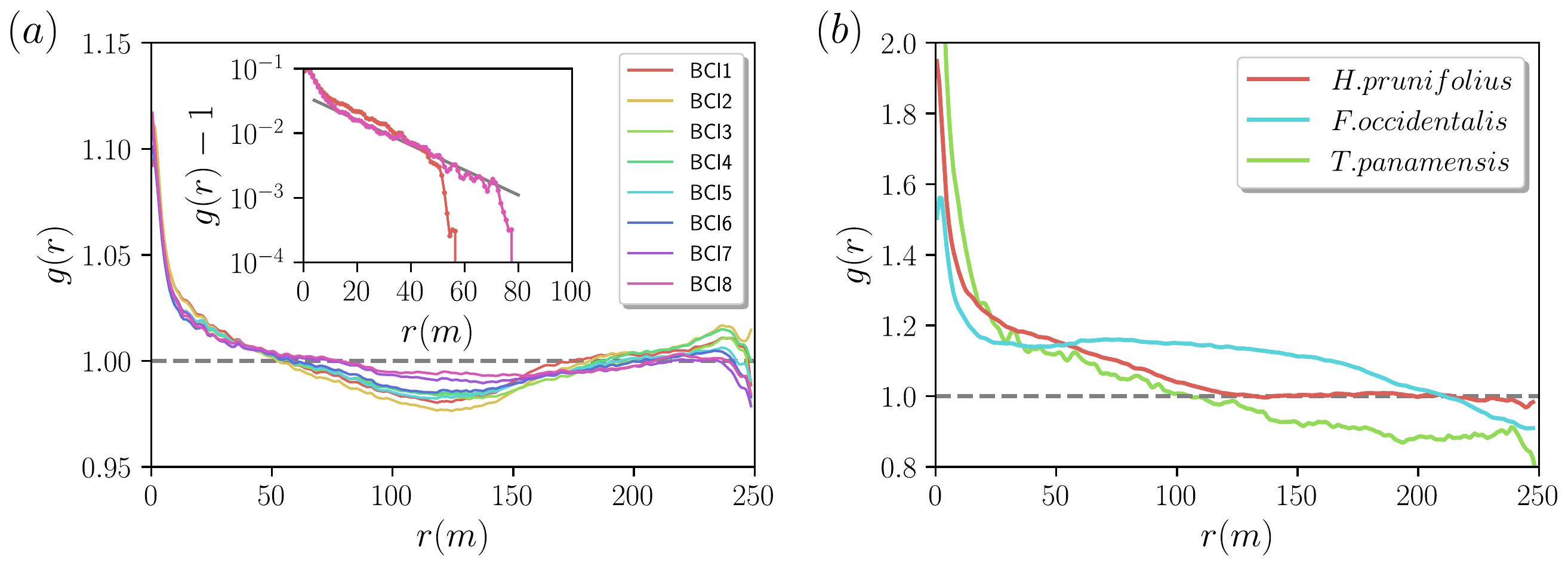}
\caption{Density correlations in BCI. (a) \new{Pair correlation function}
  $g(r)$ vs.\ distance, for the whole tree community
  and different censuses.  Inset: $g(r)-1$ in semi-log scale for the first and last
  censuses; an exponential function with correlation length $\xi=22$m
  fits the data from 10$\,$m to 70$\,$m. (b) \new{g(r)} for three 
  species selected among the ten most abundant: \emph{H.~prunifolius}
  (red), \emph{F.~occidentalis} (cyan) and \emph{T.~panamensis} (green). In both panels, the
  grey dashed line corresponds to the result for a completely
  homogeneous distribution.
  \label{FigCorr}}
\end{figure}

In Fig.~\ref{FigCorr}a we show the \new{PCF} considering all trees,
regardless of their species, and for all censuses. At small distances,
$r<50m$, $g(r)>1$ provides evidence of clumping. At larger distances,
the $g(r)$ displays a plateau to a value fairly close to 1. While
deviations from $1$ at short distances are small, they are robust as \new{evidenced}
by removing the plateau value, \new{where} a clear exponential decay with
characteristic correlation length $\xi\approx 22m$ emerges (see
inset).  Below $10 m$, deviations suggest a steeper exponential decay
though, owing to the short range of scales, precise statements are
difficult. Finally, no significant differences between various censuses
can be observed.

We now turn to the density correlations of conspecific trees. In
particular we focus on three of the more abundant species: \emph{Hybantus
prunifolius, Faramea occidentalis} and \emph{Tetragrastris
panamensis.}  For all species (see also Fig.~SI.6a), the \new{PCF} exhibits
a common behavior at short distances indicating clumping ($g(r)>1$).
Actually the deviations from $1$ at short distances are stronger than
for the community level \new{PCF}, suggesting that conspecific trees tends
to be more clustered than the whole community. However, at large
distances we found unexpected and diverse results. For
\emph{H.~prunifolius}, the \new{PCF} converges to a plateau at $1$, meaning that
at large scales the neighbour density recovers the mean density.  For
the other two species, the plateau (if any) is less well-defined, and
at values either below $1$ \emph{(T.~panamensis)} or above $1$
\emph{(F.~occidentalis)}. For other species (see Fig.~SI.6a), also
non-monotonic behaviors can be observed with clear signatures of
anticorrelation i.e. $g(r)<1$.

So far we considered as borders the edges of the rectangular plot,
\new{but this is not always the most reasonable thing to do}. The particular case of \emph{H.~prunifolius}
offers a clear-cut example. The unambiguous empty area in
Fig.~\ref{Sketch} corresponds to a swampland where \emph{H.~prunifolius}
cannot easily establish \cite{horvat2010spatially}.  Clearly, when
computing the $g(r)$, its perimeter must be considered as an internal
border. An improper identification of the borders might bias not only
the number of neighbours but also the estimation of the covered area
and thus of the mean density, leading to a wrong normalization of
$g(r)$.

For instance, \new{as exemplified in Fig.~SI.2a,} neglecting voids in
an otherwise random distribution of points leads to spurious values of
$g(r)$ above $1$ \new{(erroneously suggesting clumping)}, whereas
\new{the expected behavior $g(r)=1$ is recovered when 
  borders are correctly taken into consideration}.

\subsection{$\alpha$-shape borders}

\begin{figure}[hbtp]
\centering \includegraphics[width=0.95\columnwidth]{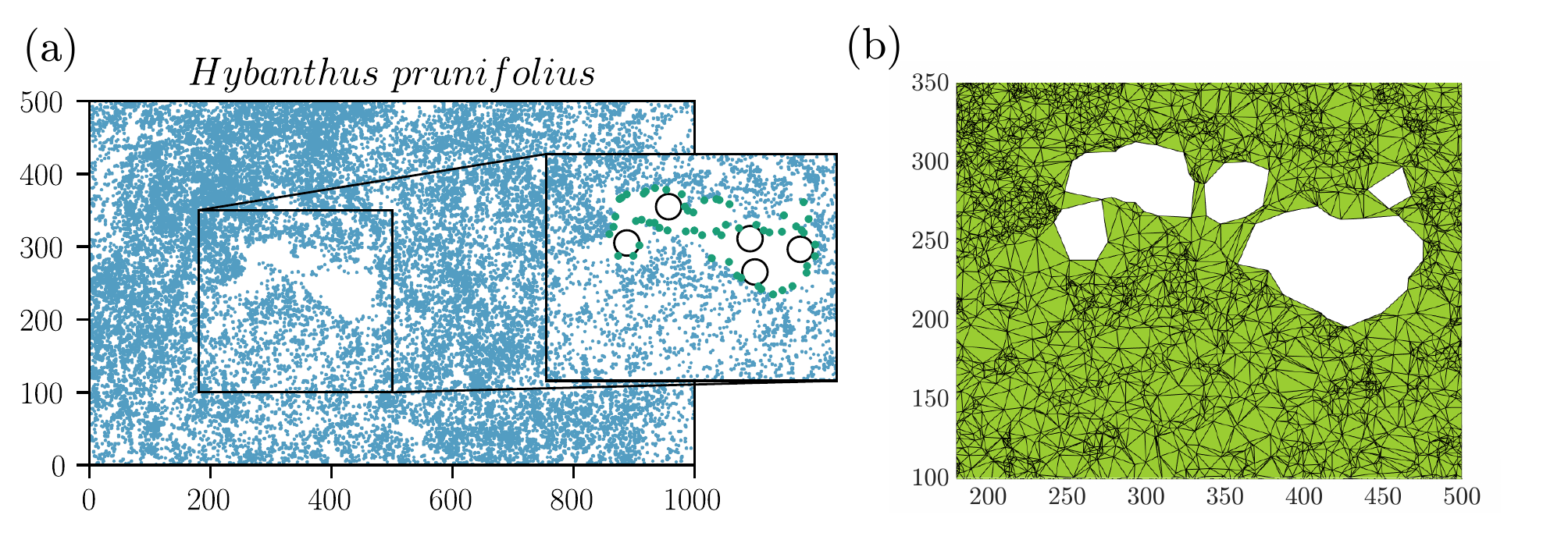}
\caption{Example of non-trivial borders in the case of
  \emph{H.~prunifolius}.  (a) The distribution of
  \emph{H.~prunifolius} individuals (points) in the 8th BCI census
  clearly displays a big empty region.  The inset shows how using
  $\alpha$-shapes (a few circles with radius $\alpha=14$ are shown)
  one can identify the internal border (green points represent the
  border trees). (b) Once the border trees are identified, Delaunay
  triangulation allows to measure the covered area, and thus
  estimate the average tree density, $\rho_\alpha$, with the empty region
  excluded.
  \label{Sketch}}
\end{figure}

\new{Finding the borders of a set of points thus becomes a crucial
  issue. A first approximation to identify non-trivial borders is to
  consider the Convex Hull (CH) of the set of points, i.e. the
  smallest convex polygon containing all the points. Nonetheless, both
  concavities and internal empty regions are key to correctly identify
  the borders.}  Here we use the $\alpha-$shapes method
\cite{Edels1983,Edels1994}, an algorithm which carves the distribution
of points under considerations with a disc of radius $\alpha$, where
$\alpha$ is a tuning parameter. \new{The border is formed by pairs of trees (points) that can be touched by an empty disc of radius alpha,
  independently of the distance between points. In other words, when a
  disc touches two points in the plane, these are added to the border if
  no other point is contained in the disc (see Fig.~\ref{Alpha} for an
  illustration and for more technical aspects see Supplemental
  material SI-1). In this way, all the concavities or voids larger
  than the radius $\alpha$ can be detected, identifying the set of
  points belonging to the non-convex border at scale
  $\alpha$. Clearly, for $\alpha$ large enough, the algorithm recovers
  the CH envelope of the system. In Fig.~\ref{Sketch}a we show an
  application of the above described method to the distribution of
  \textit{H.~prunifolius}, one can appreciate how the $\alpha$-shape
  method is able to identify the internal borders.}  Once border trees
are identified, the covered area is measured by means of Delaunay
triangulation (shown in Fig.~\ref{Sketch}b for
\textit{H.~prunifolius}) \cite{delaunay1934sphere}.

\begin{figure}[hbtp]
\centering \includegraphics[width=0.8\columnwidth]{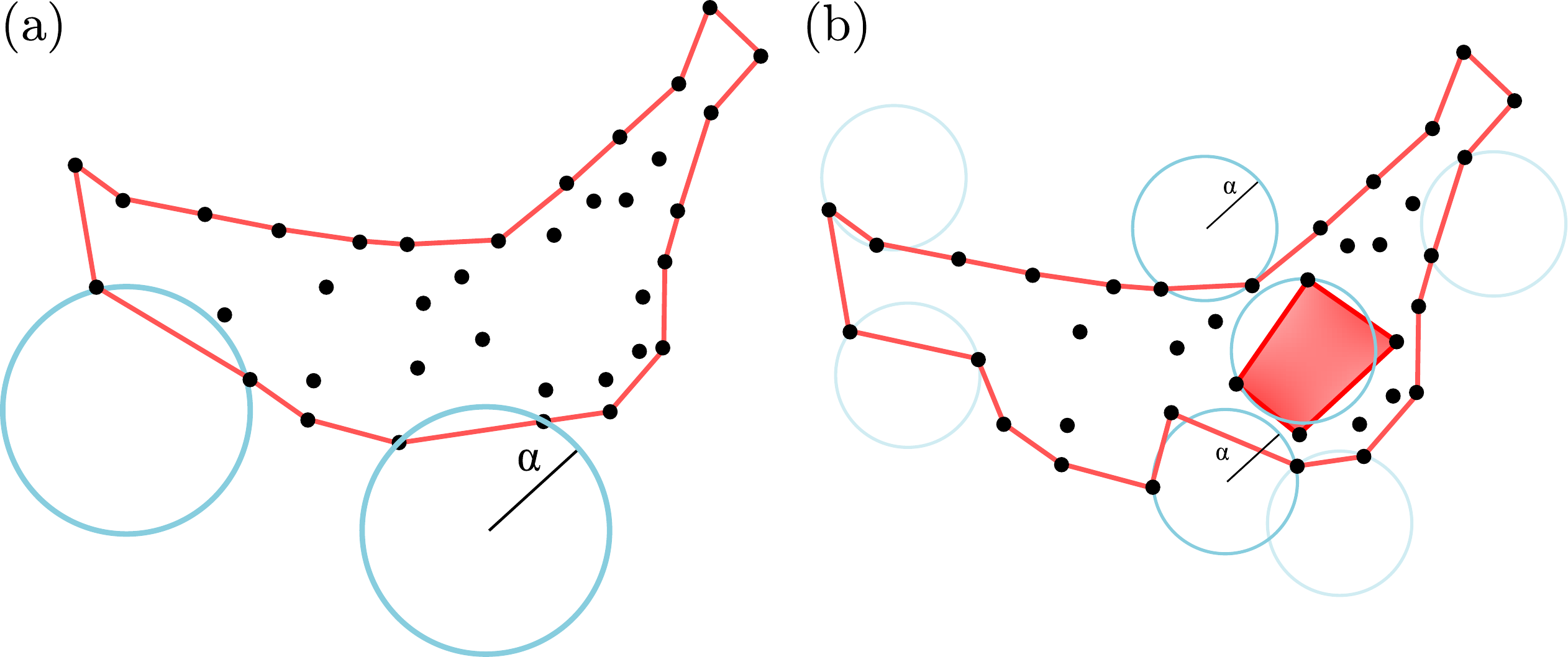}
\caption{\new{Sketch illustrating the two-dimensional $\alpha$-shape method for
 border identification. Given the set of points and a circle of radius $\alpha$, the circle is moved in the set of points. When the circle touches two points they are considered border points if there are no other points inside the circle. When the circle's radius is progressively reduced the algorithm is able to discriminate (a) external borders or (b) internal borders (red shaded area is now excluded) and concavities. }
\label{Alpha}}
\end{figure}

\new{The case of \emph{H.~prunifolius} is particularly straightforward, as the
  internal borders are easy to detect by eye and, moreover, it can
  clearly be ascribed to soil characteristics, here a
  swampland. For other species the situation is more
    ambiguous, and} one has to bear in mind that there is no rigorous
  prescription to fix the value of $\alpha$, so that the method
  involves some level of subjectivity.  In the absence of clear clues
  on the actual size of concavities, it is not obvious how to choose
  $\alpha$.  In the synthetic case discussed in Fig.~SI.2, the proper
  value can be identified by searching for a plateau of the mean
  density (or enclosed area) as a function of $\alpha$, but for the
  BCI data such plateaus are not well defined or absent (see
  Fig.~SI.3).  In selecting the value of $\alpha$, we required
  $\alpha$ to be significantly larger than the mean distance between
  neighbouring trees (to avoid artificial fragmentation) but small
  enough to ensure the identification of empty areas in the spatial
  distribution of different species. To check for the latter we looked
  for changes in the curvature of the covered area and used visual
  inspection (see discussion of Fig.~SI.3).  With the above proviso,
  we recomputed the \new{PCF}s for the three species with the borders
  identified with the $\alpha$-shapes. For the community level
  \new{PCF} the borders essentially coincide with the rectangle.
  Borders for the ten most abundant species, including the three
  species here discussed, are shown in Fig.~SI.4.

\begin{figure}[t!]
\centering \includegraphics[width=1.0\columnwidth]{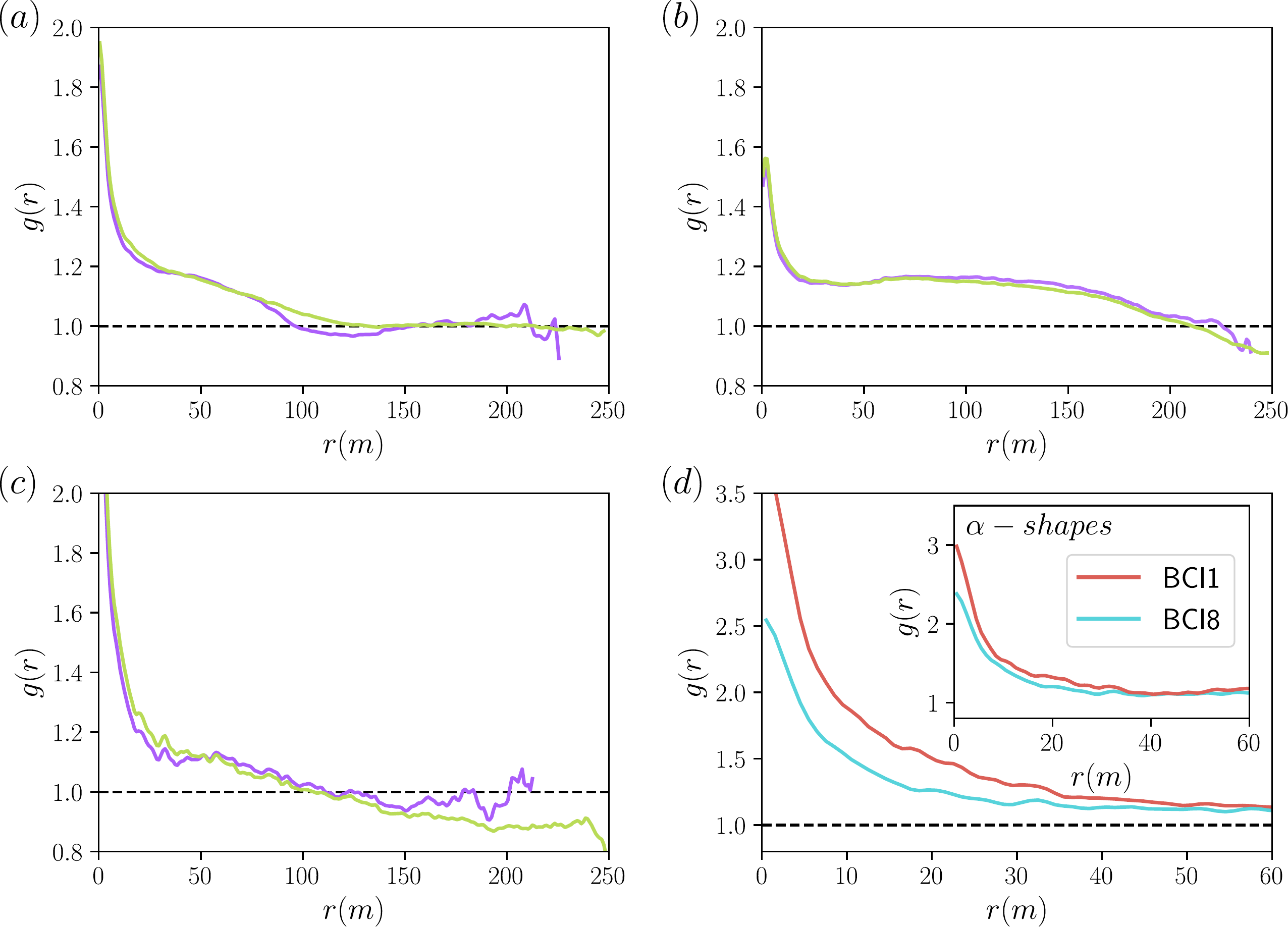}
\caption{(a-c) Comparison of the \new{pair correlation} function, $g(r)$,
  computed using the naive rectangular borders (green curves) or the
  borders identified via the $\alpha$-shape method (\new{purple} curves) the
  three selected species: (a) \emph{H.~prunifolius}, where $\alpha=14$
  allows to exclude the void shown in Fig.~\ref{Sketch}; (b)
  \emph{F.~occidentalis} for which we used $\alpha=12$; (c)
  \emph{T.~panamensis} with $\alpha=24$. For these and other species
  the identified borders are shown in Fig.~SI.4. Panel (d) shows the
  \new{g(r)} of \emph{T.~panamensis,} computed for the first (red) and last
  (cyan) censuses, with the naive rectangular borders (main panel) and
  $\alpha$-shapes with $\alpha=24$ (inset). In all panels, the black
  dashed line is the result for a uniform random distribution.
  \label{FigAlpha}}
\end{figure}

In Fig.~\ref{FigAlpha}a we compare the \new{PCF} for \emph{H.~prunifolius}
computed using both the naive rectangular border and those obtained
with the $\alpha$-shapes, which exclude the void shown in
Fig.~\ref{Sketch}.  In this particular case, $g(r)$ exhibits a
robust behavior, being mostly independent of the border
definition. Similarly, also for \emph{F.~occidentalis}
(Fig.\ref{FigAlpha}b) the dependence of \new{PCF} on the border choice is
(if any) very weak. In particular, there remains a plateau at a value different
from $1$, meaning that the neighbour density does not
recover the mean density at large scales, independently of the border definition.  In
Fig.~SI.6 we show the \new{PCF}s of the ten most abundant species
computed with the rectangular border and that given by the
$\alpha$-shapes respectively. Small scale clumping appears as a
robust feature independently of the border, thus confirming previous
findings \cite{condit2000}.

In contrast, for some species the different choice of borders has an
evident effect at intermediate and large scales.  In particular,
exclusion of empty areas through use of $\alpha$-shape leads to the
disappearance of anticorrelations ($g(r)<1$).  However, removal of
voids is a delicate issue.  Indeed, such empty areas might be the
result of some relevant ecological process, and thus their removal
could represent the introduction of an artefact \new{and thus needs to
  be checked against fair biological criteria or biogeographical
  aspects.} In this respect, the case of \emph{T.~panamensis} (which
displays anticorrelation at large distances, Fig.~\ref{FigAlpha}c-d)
is worth of attention.  In Fig.~\ref{FigAlpha}c we compare $g(r)$
computed with the naive borders and those obtained with the
$\alpha$-shapes. Here, accounting for the borders leads the \new{PCF}
to a fair plateau around $1$ at large scales while with the naive
borders it has a quasi-monotonic decay to values clearly below $1$.
The comparison between different censuses is revealing. In the main
panel of Fig.~\ref{FigAlpha}d, we show the \new{PCF} of
\emph{T.~panamensis} computed with the naive borders for census 1 and
8, showing a clear difference at small scales between the two
censuses. Conversely, the small scale behavior is basically unchanged
when considering the borders given by the $\alpha$-shapes
(inset). This behavior is explained observing that
\emph{T.~panamensis} has spread between census 1 and census 8 (see
insets of Fig.~SI.3b-d), so that the naive border overestimates the
covered area, especially in census 1.  The usefulness of using
$\alpha$-shapes for spreading species was previously highlighted in
Ref.~\cite{capinha2014predicting}.

\begin{figure}[t!]
\centering \includegraphics[width=1.0\columnwidth]{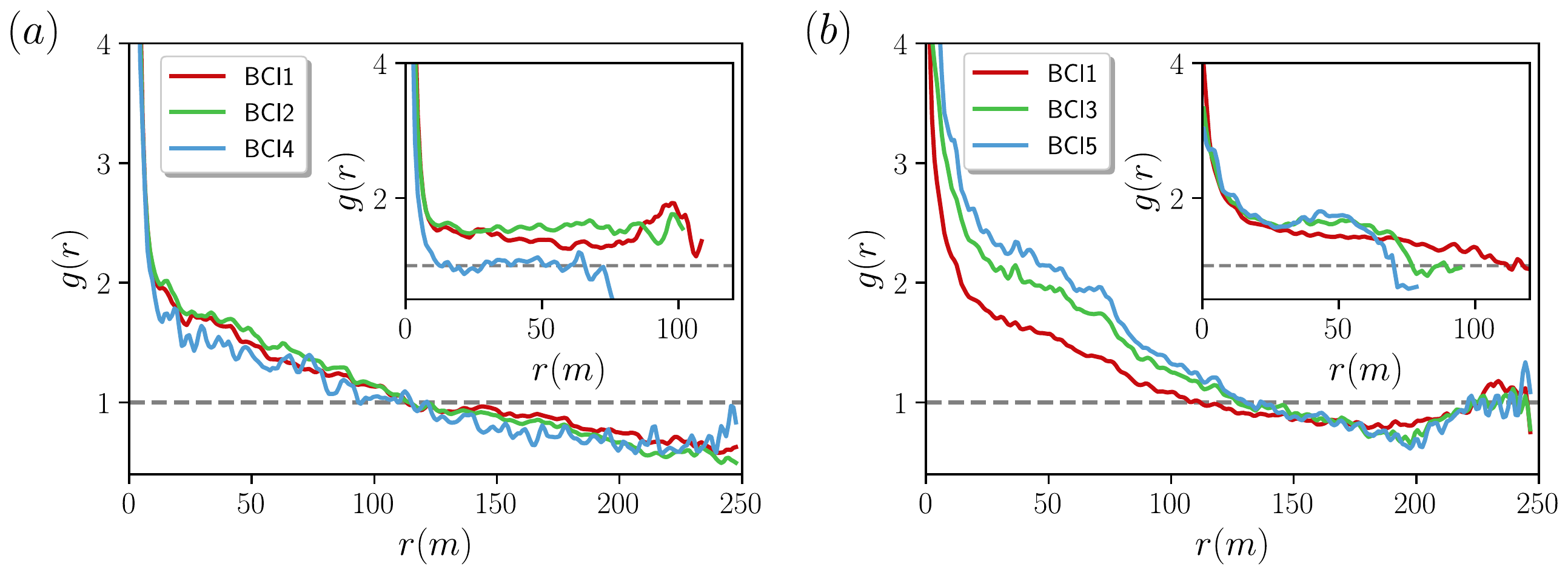}
\caption{Density correlation for contracting species.  (a) \new{Pair correlation function} for
  \emph{P.~cordulatum} computed with the rectangular border (main
  panel) and $\alpha$-shapes (inset) using $\alpha=30$ for censuses 1
  and 2, and $\alpha=32$ for census 4. (b) Same as panel (a) for
  \emph{P.~armata} and censuses $1$ ($\alpha=40$), $3$ ($\alpha=35$)
  and $5$ ($\alpha=35$).  The borders identified with the
  $\alpha$-shapes are shown in Fig.~SI.5.}
  \label{FigContracting}
\end{figure}

The choice of borders has similarly an important effect for species
contracting across censuses, as shown in Fig.~\ref{FigContracting} for
two species that experienced steep population declines. In particular,
for \emph{Piper cordulatum} one can recognise the washing out of
correlations induced by the process of disappearance of the trees
(Fig.~\ref{FigContracting}a), while for \emph{Poulsenia armata}
(Fig.~\ref{FigContracting}b) one sees that the structure of different
censuses is actually quite similar, especially at small scales,
contrary to what would result from the use of naive borders.  Visual
inspection of the tree patterns (Fig.~SI.5) also suggests that the
former, while contracting, is also loosing more features in the tree
distribution than the latter, which qualitatively explains the
difference in the density correlations.

\new{This analysis shows that, even if the PCF  --eventually with
properly defined borders-- displays a reasonable plateau at $1$ for some cases, for others we do not recover such a trait,
independently of the definition of the borders (see Fig.~SI.6)}. The
expectation of recovering $g(r)\approx 1$ at large scales essentially
relies on two assumptions: (i) at large scales the
distribution is homogeneous with a well defined (representative) mean
density; (ii) correlations have died out at the largest scales
which we can observe.  Clearly, the unmet expectations can originate by
the breaking of one or both the assumptions. These considerations
bring us to inquire about the density fluctuations, which is the
subject of the following section.

\section{Density fluctuations and Taylor's Law \label{sec:fluct}} 

For a completely random (homogeneous Poisson) process, the typical
null-model in point process analysis, density fluctuations decrease
with the square root of the area over which the density itself is
estimated.  It is instructive to see how this is achieved. Given $N$
points in an area $A$, the sample mean density is $\rho_0=N/A$. Divide
the area $A$ in cells, e.g.  squares of side $r$, and denote with
$n_r(\bm x)$ the number of points in the cell centered in $\bm x$, by
definition $\rho_0=\langle n_r(\bm x)\rangle/r^2$ where
$\langle [\dots]\rangle$ indicates the average over all cells. To
study density fluctuations, we first define the coarse-grained local
density $\rho_r(\bm x)=n_r(\bm x)/r^2$ at scale $r$ and then look at
its root mean square deviations normalized by the mean density,
\begin{equation}
  \frac{\delta_r \rho}{\rho_0}\equiv \frac{\langle (\rho_r(\bm x)-\rho_0)^2\rangle^2}{\rho_0}=
  \frac{[\langle n^2_r(\bm x)\rangle-\langle
      n_r(\bm x))\rangle^2]^{1/2}}{\rho_0r^2}\equiv \frac{\delta_r n}{\langle n_r\rangle}\,,
\label{eq:bridge}
\end{equation}
where in the last equality we used that $\langle n_r(\bm
x)\rangle=\rho_0 r^2$, and dropped the dependence on $\bm x$, for
simplicity.  For a homogeneous Poisson process, $(\delta_r n)^2=
\langle n_r\rangle$, and Eq.~(\ref{eq:bridge}) implies that
fluctuations decay with the square root of the sampled area,
$(\delta_r \rho/\rho_0) = \langle
n_r\rangle^{-1/2}=\rho_0^{-1/2}r^{-1}$.

However, it is an empirical
observation that for many ecological processes
\begin{equation}
\delta_r n\propto \langle n_r\rangle^{\gamma}
\label{eq:TL}
\end{equation}
with an exponent $\gamma$ typically ranging in $1/2\leq \gamma \leq
1$. Using Eq.~(\ref{eq:TL}) with $\langle n_r\rangle=\rho_0 r^2$ yields 
\begin{equation}
  \frac{\delta_r \rho}{\rho_0} \propto r^{2(\gamma-1)}
\label{eq:fluct}
\end{equation}
for density fluctuations. Consequently, when $\gamma>1/2$, and the
more it approaches $1$, such fluctuations become more and more
important and decrease with the observation scale much slower than for
a random homogeneous process.

\begin{figure}[hbtp]
  \centering
\includegraphics[width=0.8\columnwidth]{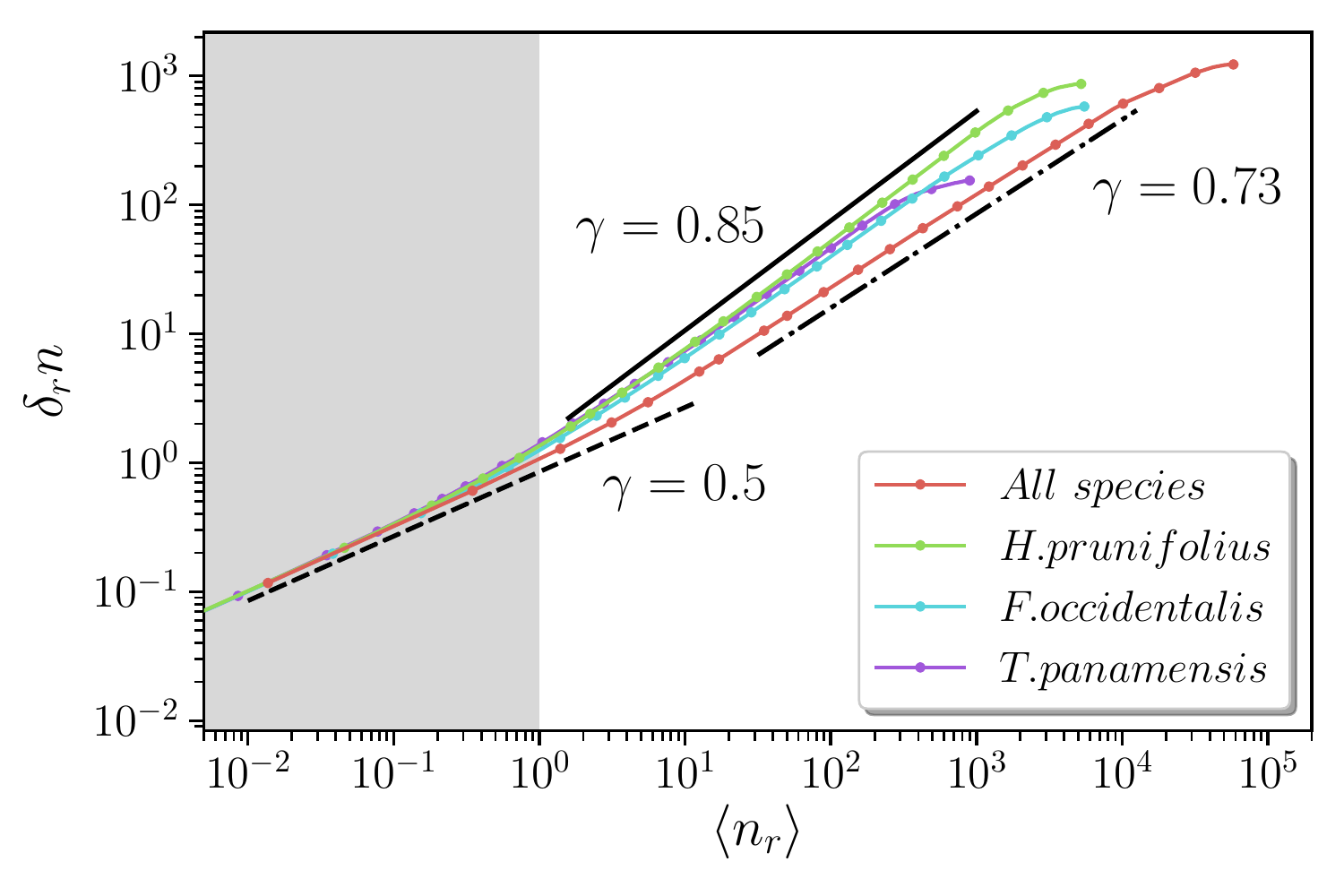}
\caption{\new{Spatial} density fluctuations: standard deviation of number of trees,
  $\delta_r n=[\langle n^2_r\rangle-\langle n_r)\rangle^2]^{1/2}$, as
  a function of the mean number of trees, $\langle
  n_r\rangle$. Borders are identified via $\alpha$-shapes,
  and bias is cured as described in the text. Lines
  display the Taylor's law exponents $\gamma=1/2$ (dashed,
  corresponing to a homogeneous random process), $\gamma \approx 0.85$
  (solid, the value that describes the labeled species with some
  variability of the order of $0.03$),  and $\gamma \approx 0.73$ (dot
  dashed, describing the community behaviour).
\label{GDF}}
\end{figure}

The power law behavior (\ref{eq:TL}) is known in the literature as
Taylor's Law (TL) and it was first put forward in the context of
population ecology \cite{taylor1961} (see Ref.~\cite{Eisler2008} for a
review). The empirical relationship (\ref{eq:TL}) is found both in
spatial distribution and in the temporal evolution of biological
populations, from trees to birds and insects
\cite{anderson1982variability, reed2004relationship}, as well as a
wide variety of systems including stock markets, heavy-ion collisions,
traffic in complex networks, population ecology \cite{botet2001,
  arenas2006, kilpatrick2003} and active matter, where it is known as
giant density fluctuations \cite{chate2006simple, narayan2007}.
Dependencies between individuals or environmental variability have been
invoked to rationalize the ubiquitous emergence of such scaling law in
ecology \cite{kilpatrick2003,giometto2015, james2018}.
Ref.~\cite{Eisler2008} thoroughly reviews the possible mechanisms put
forward to explain a deviation from $\gamma=1/2$, which as shown above
corresponds to the random (Poisson) distribution and, more generally,
to cases in which the central limit theorem applies.  Values of
$\gamma$ in the interval $(1/2,1)$ are typically indicative of long
correlations, a hallmark of out-of-equilibrium systems, and/or the
presence of \new{spatial} heterogeneities \cite{Eisler2008}.

Similarly to the \new{pair correlation} function, however, the presence
of borders must be properly taken into account to avoid wrong
estimation of density fluctuations (Fig.~SI.2b). To cure border
induced biases, we adapted the Hanisch method \cite{hanisch1984some},
previously discussed for the \new{PCF}. A large number of random points is
drawn in the rectangular BCI plot. Each point is retained only if it
is not outside or on the borders defined by the $\alpha$-shapes, then
for each valid point $i$ one measures the distance $d_i$ from its
closest border and then compute the number of trees contained in
circles centered in $i$ and of radius $r<d_i$. Averaging over all
points (whose number should be large enough) and considering different
radii, one can estimate $\langle n_r\rangle$ and $\delta_r n$. In this
case $\langle n_r\rangle=\pi \rho_\alpha r^2$ as the density is
estimated dividing the number of trees with the actually covered area
for that value of $\alpha$ and $\pi$ accounts for using
circles\footnote{Of course one could also use square cells, but
  circles are more practical in the presence of non trivial borders.}.
Apart from this inessential change, the above description of the link
between TL and density fluctuations remains unchanged.

Figure~\ref{GDF} shows that data from the BCI plot obey 
Taylor's law (\ref{eq:TL}) with exponent $\gamma$ significantly
greater that $1/2$ both at the community and single species levels,
except at the smallest scales (shaded area of Fig.~\ref{GDF}, where
$\langle n_r\rangle<1$).  However, $\gamma=1/2$ at short scales is not
due to the recovery of randomness, but to the fact that for scales
much smaller than the typical distance between trees, most of the
cells are empty, and a few contain a single tree, so that
$\langle n_r^2 \rangle=\langle n_r\rangle\ll1$ and trivially
$\gamma=1/2$ \cite{Eisler2008}. \new{More in general, for very small distances,
  the second moment is dominated by fluctuations statistics that must be Poissonian as previously observed in Ref.~\cite{seri2015spatial}.}

At intermediate and large scales, for
the previously examined \emph{H.~prunifolius}, \emph{F.~occidentalis}
and \emph{T.~panamensis,} we find $\gamma$ is compatible with 0.85,
while the exponent is close to 0.73 for the whole community, revealing
the possible influence of heterogeneities and/or long range
correlations.

Looking at the ten most abundant species (Fig.~SI.7), we find that
$\gamma$ is always greater than $1/2$, and ranges in the interval
$[0.75, 0.85]$, \new{in fairly good agreement with previous studies
  \cite{seri2015spatial}.}  The same figures shows that, for the most
abundant species, the effect of borders is quite small on the TL.  But
the situation is different for the contracting species that we
discussed in Fig.~\ref{FigContracting}, where the change in $\gamma$
(when using properly defined borders) is quantitatively more important
(see Figs.~SI.8 and SI.9). In particular, data for
\emph{P.~cordulatum} on census 4 display the strongest difference,
approaching values of $\gamma$ close to $1/2$, thus confirming the
tendency toward recovering a homogeneous random distribution suggested
by the density correlations.

From the above analysis we can conclude that density fluctuations are
always anomalous ($\gamma>1/2$) for the whole BCI community and for
most abundant single species.  Such anomalous (with respect to the
completely random process) density fluctuations observed at large
scales constitute an indicator of high spatial heterogeneity and/or
long range correlations.

\section{Theoretical considerations on modeling the data \label{sec:models}}

In this section, with the aid of some illustrative reference models,
we discuss the usefulness of combined information from density
correlations (the \new{PCF}) and density fluctuations (Taylor's law) when
interpreting and modeling real data.  In a growing order of
complexity, we will consider patterns obtained with a very simple
Heterogeneous Poisson Process (HPP), the Thomas Process (TP), and a
spatially explicit neutral model. These models are used only for
testing how some general mechanisms influence density correlations and
fluctuations, and are not proposed as explanatory models for BCI
data. However, for the sake of qualitative comparison,
for the HPP and TP we have chosen a rectangular domain
$1000\times 500$, similar to BCI plot, and a number of trees of the
order of the most abundant species in BCI. For the neutral model, due
to the characteristics of the model, a different criterion has been
chosen.

\subsection{Heterogeneous Poisson Process}

As discussed in Sec.~\ref{sec:fluct}, one of the possible origin of a
Taylor exponent greater than $1/2$ (characteristic of the homogeneous completely
random process) is the presence of inhomogeneities.  To
illustrate this point we consider a very simple heterogeneous Poisson
process, with non uniform intensity. In particular, the
$1000\times 500$ domain is divided in two halves, each one characterized by
a uniform density $\rho_1$ and $\rho_2$. In particular,  we take $N=3\cdot10^4$ points
placing $N_1=3N/4$ and $N_2=N/4$ of them in the first and second half
of the rectangular domain, respectively.

Figure \ref{Models}a shows a realization of the process (left) with
the \new{PCF} (middle) and TL (right), computed on that instance. Notice
that the $g(r)>1$ as it is dominated by the most abundant points
having density larger than the mean, which is used for
normalization. A closer inspection in fact reveals that at small
distances $g(r)$ converges to the weighted average density
$\rho_w=(N_1\rho_1+N_2\rho_2)/N$ normalized by $\rho_0$. This simple
example, demonstrates that using $g(r)>1$ as the sole
criterion for clumping can be misleading. Here the visual inspection
of the point patterns confirms that it is not clumping that leads to $g(r)>1$.

In general, in the presence of heterogeneities, one of the main
problems is the normalization with the mean density $\rho_0$, which
does not represent the true density in any region of the space.  This
is clearly demonstrated by looking at the density fluctuations. The
latter displays the trivial $\gamma=1/2$ at very small scales, where
$\langle n_r\rangle <1$. While, at the interesting scales, it shows a
clear power law behavior with $\gamma=1$, meaning that density
fluctuations remain constant over the scales, see
Eq.~(\ref{eq:fluct}). This looks trivial given the way the patterns
has been generated, however, values of $\gamma\approx 1$ can be
observed also for non-trivial reasons
\cite{chate2006simple,narayan2007,Eisler2008}.

This example is deliberately oversimplified, in natural
point process the density field is not expected to vary like a step
function. Less trivial heterogeneities may lead to exponents
$1/2<\gamma<1$, and the $g(r)$ may be more complicated.

\subsection{Thomas Process}

Small scales clumping appears to be a robust feature of rainforests
\cite{condit2000}.  As the HPP example shows, heterogeneities may lead to
$g(r)>1$ and, from an ecological point of view, this may be due to the presence of more
favourable terrain, enhancing the chances of establishment and/or the
number of offspring dispersed by an adult tree. However, seed dispersal
limitation alone may generate clustering \cite{condit2000}.
Here, we consider \new{the simplest kind of point process able to mimic such clustering mechanisms, i.e. those} belonging
to the class of Poisson cluster processes \cite{Stoyan1994,MWBook} \new{and, in particular}
 the Thomas Process (TP) \cite{thomas1949}. TP \new{assumes that}: i) $n_p$ \new{original} centers are distributed randomly
according to a homogeneous Poisson process with density $\rho_p$, ii)
each \new{center} generates ---following a Poisson distribution--- $\mu$
offspring, \new{deployed with} a Gaussian kernel
\new{around the central} tree with standard deviation $\sigma$, \new{mimicking the dispersal distance of seeds}.

\begin{figure}[hbtp]
\centering \includegraphics[width=1.0\columnwidth]{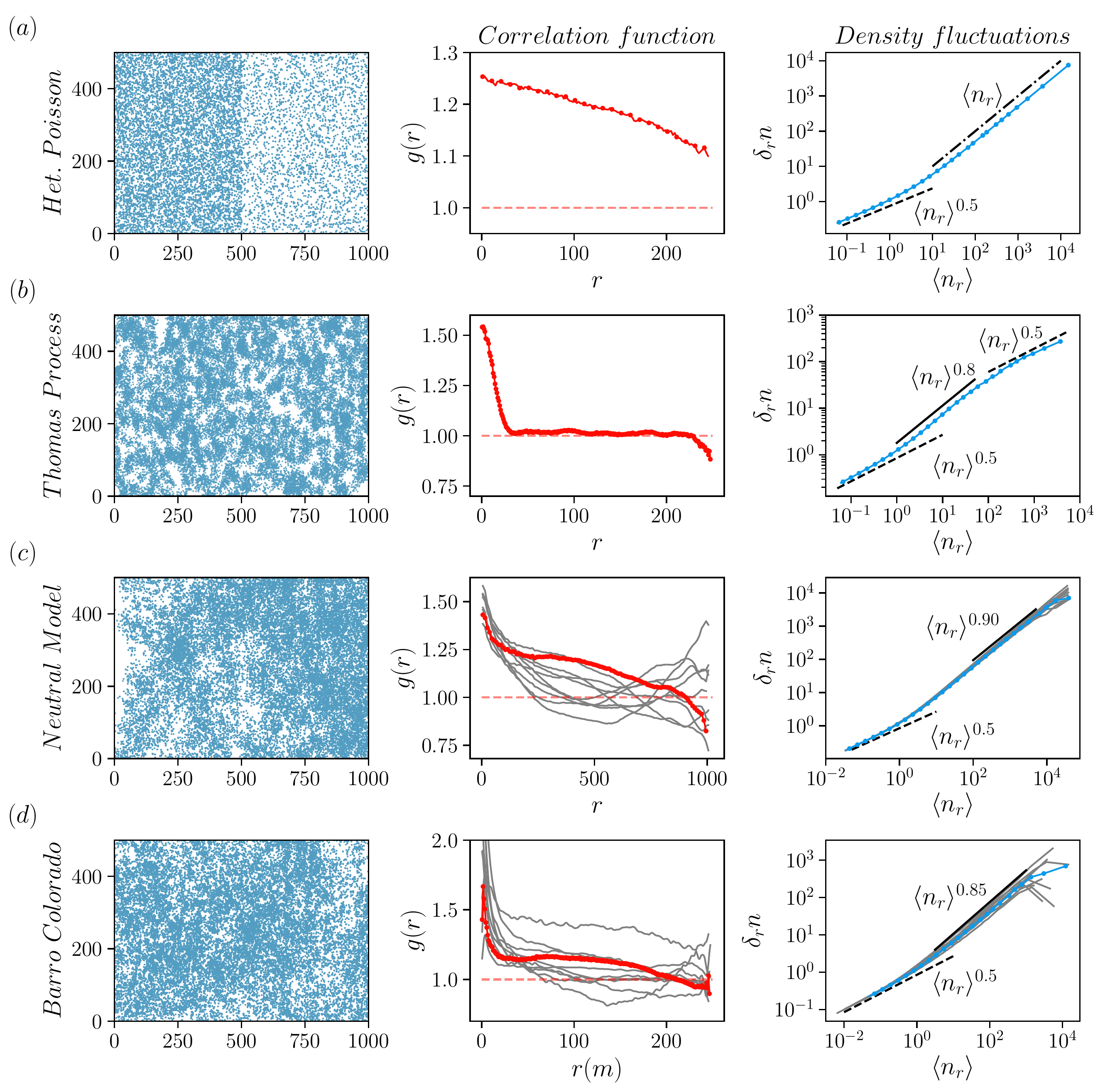}
\vspace{-0.7truecm}
\caption{Spatial pattern (left) and the corresponding \new{PCF} (centre) and
  Taylor's law (right) for the cases of (a) the heterogeneous Poisson
  process with $N=N_1+N_2=3\cdot10^4$ points $N_1=3N/4$ and $N_2=N/4$
  distributed on the two halves of the $1000\times500$ rectangle; (b)
  the Thomas process with $n_p=10^3$ adult trees each spreading
  $\mu=30$ offsprings with a Gaussian kernel with $\sigma=9$; (c) the
  multispecies voter model simulated in a $2048\times2048$ lattice with
  $\nu=3.8\cdot10^{-6}$ and $\sigma=9$, for which we selected 10
  species among the most abundant ones (with $N\approx 1.4$--$1.6\cdot10^5$
  trees) to compute the \new{PCF} and TL. For the spatial
  pattern we shown only a $1000\times500$ rectangular portion, for an easier
  qualitative comparison with the other figures; (d) BCI data: the
  spatial pattern is for \emph{Faramea occidentalis} and the \new{PCF}
  and TL are shown for the ten most abundant species. In (c) and (d), red/blue curves
  refer to \new{PCF}/TL computed on the displayed pattern, while 
  grey lines refer the rest of the ten most abundant species.  Dashed curves
  display theoretical expectations for a homogeneous process for \new{PCF}
  (red) and TL (black). The dash-dotted line in (a) shows
  $\gamma=1$ for TL, while the solid lines shows a reference slope for
  the TL exponent.
  \label{Models}}
\vspace{-0.7truecm}
\end{figure}

Figure \ref{Models}b shows a single realization with $n_p=1000$,
$\mu=30$ and $\sigma=9$ (chosen to have the small scales density of
neighbours in the range of values observed in the selected species of
the BCI plot), with associated \new{PCF} and TL.  The \new{PCF} displays clear
signs of short range clumping and a plateau at $g(r)=1$ for large scales,
meaning that correlations die out and the density of neighbours
converges to the mean density $\rho_0$. The TP is widely employed in
the ecological literature, as the \new{PCF} has a simple Gaussian-like
analytical expression \cite{Stoyan1994,MWBook},
$g(r)=1+\exp[-r^2/(4\sigma^2)]/(4\pi\sigma^2\rho_p)$, which can
be used for fitting data \cite{wiegand2009,leithead2009}. From the
above expression one can readily see that the deviation from $1$, and
thus the intensity of clustering, is controlled by both the dispersal
distance ($\sigma$) and parents density ($\rho_p$). At large scales,
owing to the random distribution of the \new{primary trees}, the process recovers
a homogeneous distribution and thus a well defined mean density. This
can be appreciated, even without knowing how the process was
generated, by looking at the density fluctuations. At large scales
indeed, the TL exponent $\gamma$ approaches the random distribution
value $1/2$.  At shorter scales, but above the inter-particle distance (marking the
end of the trivial short-range $\gamma=1/2$ regime), an exponent
$\gamma=0.8$, incidentally close to the typical values of BCI species,
can be observed. At these scales correlations, clumping and associated
inhomogeneities in the density of points are at play, leading to
$\gamma>1/2$. It is interesting to contrast the behavior of the
density correlation and fluctuations with \emph{H.~prunifolius}.  There,
while the $g(r)$ was reaching a plateau compatible with $1$ at large
scales (Fig.~\ref{FigAlpha}a), suggesting that correlations have died
out, the density fluctuations (characterized by $\gamma\approx 0.85$
at all available scales) show a slower recovery of homogeneity with
respect to the random process. This information could have not been
obtained looking only at the \new{PCF}.


\subsection{Spatially explicit neutral model}
We now study single species patterns generated by the multispecies
voter model (MVM), a spatially explicit neutral model
\cite{Durrett1996,pigolotti2018}. Here, statistical properties are not
preassigned as in the previous examples. They emerge from the
underlying processes: dispersal limitation, demographic fluctuations
and competition. Within the neutral framework, species are equivalent
at the individual level: their birth/death rates and dispersal
mechanism are the same and all compete for space \cite{hubbell2001}.
The MVM \cite{Durrett1996} incorporates such ideas as
follows. Consider a square lattice of size $N=L^2$, in which each site
is always occupied by a tree. At each time step a random tree dies and
is replaced: with probability $(1-\nu)$, by a copy of a random tree in
its neighbourhood (dispersal); with probability $\nu$, by a tree of a
brand-new species (speciation). The neighbourhood is defined via a
dispersal kernel (here a Gaussian with standard deviation $\sigma$).
Provided the dispersal length is finite and \new{comfortably} larger than the
lattice spacing, the kernel functional shape is not too important
\cite{rosindell2007species}.  Within this model species appear by
speciation, grow and disappear due to demographic stochasticity and
competition (controlled by local abundances) with other species,
generating a (non-equilibrium) stationary
state with the number of species fluctuating around a mean value, fixed
by $\nu$ and $\sigma$.  We numerically generated patterns at such
stationary state exploiting the duality of MVM with a system of
coalescing random walkers with an annihilation rate \cite{Durrett1996,
  Bramson1996} (see Sec.~SI-3 and \cite{pigolotti2018} for details).

Given the model characteristics, we cannot fix a priori the number of
trees or the covered area.  Moreover, the generated point patterns
will depend in a non-trivial way on parameters $\nu$ and $\sigma$, whose
systematic study, though interesting, is out the scope of this work.
For qualitative comparison, we required the relative rank abundance
obtained with the MVM to mimic that of the BCI plot (Fig.~SI.10), and
the small scale density of neighbours of the most abundant species to be in the range of values observed in BCI data. With these two criteria we could fix
$\nu=3.8\cdot10^{-6}$ and $\sigma=9$, \new{highlighting that} small variations around these
values do not change the results.

In Fig.~\ref{Models}c, we display a typical spatial pattern generated
with the MVM. We also show the density correlations and \new{spatial} fluctuations
for several species in the same class of abundance.  Besides small scales clumping \new{is a} common \new{feature for} all species, we
observe diverse behaviors (with plateaus at $1$ or different from
$1$ and also non-monotonic behaviors) at large distances, in
remarkable qualitative agreement with those observed in the most
abundant BCI species, shown in Fig. \ref{Models}d. \new{We have also checked that
exclusion of empty areas through the use of $\alpha$-shape only leads to the
disappearance of anticorrelations ($g(r)<1$, see SI-3.1).  However, in this specific case the ``ecosystem'' is homogeneous by construction and empty areas are just a dynamical effect arising from neutral competition and dispersal limitation. Thus, there is no reason --certainly not from the ecological and geographical environment-- to remove such empty areas.} Moreover, as shown
in the right panels of Fig.~\ref{Models}c and d, the density
fluctuations of the patterns generated by the neutral model display,
at all available scales (above that for which $\langle
n_r\rangle\approx 1$), a non-trivial scaling behavior with an exponent
$\gamma$ in the range $0.85-0.90$, \new{compatible with those observed in} BCI data where $\gamma \approx (0.75-0.85)$.

The MVM is, by construction, spatially homogeneous, as each site of
the lattice is equivalent to the others, therefore the variability
observed in the \new{PCF} cannot be attributed to (extrinsic) spatial
heterogeneity. On the other hand, the behavior of the density
fluctuations --\new{which we have verified to be unsensitive to the inclusion of $\alpha-shapes$ method in the neutral model (see SI-3.1)}-- points in the direction of a highly heterogeneous process
as witnessed by \new{the anomalous value of} $\gamma$. This is confirmed by the
visual inspection of the point process (left panel of
Fig.~\ref{Models}c and Fig.~SI.11) with, possibly, persistent
correlations in spite of the rather short (with respect to the system
size) dispersal distance.

\section{Conclusion\label{sec:end}}
We studied spatial tree patterns by analysing the scale dependence of
density correlations, probed via the \new{pair correlation} function, and
of \new{spatial} density fluctuations, which is tightly linked to the Taylor's power
law. In particular, these tools were employed to study several species
of the Barro Colorado Island plot \cite{bcidata}. We showed that, in
order to properly estimate both observables avoiding spurious
behaviours, the borders of the census plot must be treated carefully.
Properly dealing with borders entails two issues.  The first is
avoiding the biases introduced by the points near the border, which we
solved by the Hanisch method \cite{hanisch1984some}. The second, and
more delicate, is to identify the borders. We showed how the
$\alpha$-shapes algorithm \cite{Edels1983,Edels1994} can serve to such
purpose, though with some unavoidable level of subjectivity.  The
$\alpha$-shapes revealed to be particularly important when analyzing
expanding (as e.g. \emph{T.~panamensis}) or contracting
(e.g. \emph{P.~cordulatum} or \emph{P.~armata}) species across
different censuses (see also \cite{capinha2014predicting} for previous
observations in this regard).

The small scale behavior of the density correlations confirmed the
prevalence of tree clumping \cite{condit2000}. \new{Also, we} found that
conspecific trees are generically more clustered that the whole
community. For expanding and contracting species, clumping intensity
does not seem to depend much on census (provided borders are properly
identified), with the exception of \emph{P.~cordulatum}, for which density
correlations and fluctuations suggest a tendency toward increased
homogeneity and loss of correlations with decreasing abundance.
Conversely, the large scale behavior of density correlations is much
more complex. For some species a plateau fairly close to the expected
value of a homogeneous random process was found, while for others the
plateau was at different values (non monotonic behaviours were
also found, but they tend to disappear when $\alpha$-shape borders are
implemented). These hard to interpret results were partially clarified
by the analysis of density fluctuations, whose scale decay was shown
to be related to the exponent of Taylor's law. For most
species we found that the Taylor's power law exponent is larger than
1/2 (the value of a homogeneous process), suggesting the
presence of spatial heterogeneities \new{and/or
long range correlations}.

The usefulness of the joint assessment of density correlations and
fluctuations was further demonstrated by analyzing them in three
models for point patterns, which also served for a qualitative
comparison with the field data. In particular, with a very simple
heterogeneous Poisson process we exemplified how inhomogeneities leads
to density correlations typical of clumped distributions \new{at mid-scales} and \new{trivial} strong
 \new{spatial} density fluctuations. We then examined the Thomas Process, belonging
to the class of Poisson cluster processes which statistically mimic
the dispersal of offsprings by adult trees, which are randomly and
homogeneously distributed. Here, the correlation function exhibits
clumping at short scales and, for the chosen parameter values, \new{spatial} density
fluctuations are characterised by two power laws: one with anomalous
exponent at intermediate \new{scales} (due to the correlations and inhomogeneities
caused by the clusters of offsprings), and one with exponent 1/2 at
large scales (corresponding to the homogeneous random process
controlling the distribution of adult trees).  This is different from
what is observed in field data.  Although one could probably better
mimic the empirical data with different choices of the parameters and
more \textit{ad hoc} clustering models --e.g.\ drawing the parent
trees with a more complicated processes, \new{like a double-cluster process for $F.occidentalis$ \cite{wiegand2009}}, or using different dispersal
kernels-- it is not obvious whether it is possible to fit
simultaneously both the density correlations and fluctuations.

The above models with preassigned statistical features may surely
serve as a ``fitting models'', or to illustrate a particular effect,
but it is doubtful that they can be useful as ``explaining models''.
In this respect we think that the use of individual-based models, in
which the statistical properties of the point patterns are not imposed
but arise as a result of local dynamical rules, can be more
interesting. Indeed individual-based models may present a theoretical
playground to build more stringent strategies for inferring the
ecological process from the generated pattern, which can then be
implemented in real data analysis.  This view is partially motivated
by the ability of the spatially explicit neutral model we studied to
produce patterns characterized by \new{spatial} density correlations and
fluctuations in qualitative agreement with field data.

This program will probably require introducing other tools and
observables, besides density correlations and
fluctuations. Interesting steps in this direction have been undertaken
e.g.\ using wavelets \cite{detto2013fitting}, although applied to
single species.  Spatially explicit neutral models or stochastic niche
models \cite{tilman2004niche} are able to simulate entire communities
with known rules, adding effects due to competition or heterogeneities
which are surely playing a pivotal role in ecological patterns and are
difficult to be inferred. In this respect it is surprising that a
simple neutral model with the (possibly unrealistic) assumption of
species equivalence can not only reproduce macro-ecological
biodiversity patterns \cite{azaele2016statistical,pigolotti2018} but
also single species tree patterns, at least qualitatively. This
suggests that the neutral model can be used as a null model against
which to compare tree patterns also at the single species level.

We hope our work will stimulate the study of point patterns generated by
neutral or niche spatially explicit models, which can lead to designing
better observables and tools for understanding the ecological processes
underlying the observed patterns.

\section*{Acknowledgment}
We thank M.A. Mu\~noz and S. Pigolotti for very useful comments. This
work was supported by ERC grant RG.BIO (Grant No. 785932) to A.C., and
ERANet/LAC grant CRIB (project ELAC2015/T01-0593) to A.C., H.F. and
T.S.G.. T.S.G. also acknowledges support from CONICET,
ANPCyT, and UNLP (Argentina).

\section*{Author Contributions}
A.C., T.S.G., H.F. and M.C. designed the research, P.V. performed the
research, analysed the field and numerical data.  All authors
discussed results.  P.V. and M.C. wrote the paper with input and
revision from all the authors.


\bibliographystyle{vancouver.bst}
\def\url#1{}



\section*{Supplementary material}

# Supplementary Information: Joint assessment of density correlations and fluctuations for analysing spatial tree patterns

**Villegas, P.**[1,*], **Cavagna, A.**[1,2], **Cencini, M.**[1], **Fort, H.**[3], **and Grigera, T.S.**[1,4,5,6]

[1]Istituto dei Sistemi Complessi, Consiglio Nazionale delle Ricerche, via dei Taurini 19, 00185 Rome, Italy
[2]Dipartimento di Fisica, Università Sapienza, 00185 Rome, Italy
[3]Institute of Physics, Faculty of Science, Universidad de la República, Iguá 4225, Montevideo 11400 Uruguay
[4]Instituto de Física de Líquidos y Sistemas Biológicos — CONICET and Universidad Nacional de La Plata, La Plata, Argentina
[5]CCT CONICET La Plata, Consejo Nacional de Investigaciones Científicas y Técnicas, Argentina
[6]Departamento de Física, Facultad de Ciencias Exactas, Universidad Nacional de La Plata, Argentina
[*]Correspondence and requests for materials should be addressed to pvillegas@ugr.es

## SI-1: Identification of the border with alpha shapes

As discussed in the main text, the statistical estimators able to avoid the biases induced by the borders can only be applied once these have been identified. Finding the borders of a set of points is thus a crucial issue. In the main text we discuss a powerful method to define the (internal and external) perimeter of a set of points.

A first approximation to the borders is to consider the Convex Hull (CH) of the set of points. With simple algorithms one can find the list of points conforming the convex envelope of the system. For most species in the BCI census, this essentially corresponds with the edges of the rectangular area of the plot. While finding the CH of a set of points is typically fast, field data are in general not convex. A clear example of such a problem consists of an hypothetical spatial distribution of trees surrounding an empty area, e.g. the forest glade exhibited by *H. prunifolius* (Fig. 2 of main text). In a case like this, besides the biases on neighbour counting induced by the presence of the void space, the CH method overestimates the covered area, and consequently underestimates the mean density.

As explained in the main text, a better determination of the borders can be achieved using the so-called $\alpha-$shape method (AS). This procedure is able to delineate both the internal and external edges – which are neither necessarily convex nor connected– of the system by filling it with discs of radius $\alpha$ (a tuning parameter of the algorithm, which in general is not possible to fix in an objective way).

Border points are identified by considering all point pairs that can be touched by an empty disc of radius alpha. Schematically, the algorithm works as follows: (i) Once two points have been selected (black circles in Fig. SI.1), two circles of radius $\alpha$ are centered at each point, calculating the intersection points of the two circles (red points in Fig. SI.1). (ii) From the two intersection points of the circles, draw two new circles of radius $\alpha$ (red circles in Fig. SI.1), which by construction contain the black points. (iii) If there are no points inside one of the circles, black points are considered border points. Specifically,

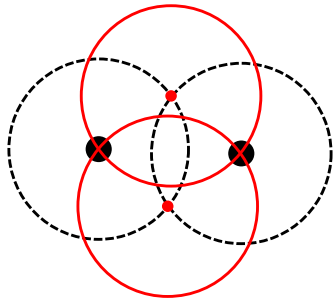

**Figure SI.1.** Two discs 'hitting' the black points can be plotted. If either of the new circles contains no points in one of such red discs, black points are considered like border points.

to compute the borders with $\alpha$-shape we used a dedicated and efficient Matlab library,[†] and computed the covered area via Delaunay triangulation, which is also implemented in the same library.

The remaining problem is now the choice of the free parameter $\alpha$. If the value of $\alpha$ is too large we get back the convex hull, while if it is too small, all the discs will penetrate into the distribution of points, breaking it up completely and producing a collection of small sets of points. Given these limiting cases, typically a good way of finding an appropriate value of $\alpha$ is to study the behaviour of the covered area (or, equivalently, of the density) as a function of $\alpha$. Thus, when the concavities of different sizes are eliminated, a sudden jump of the area is expected, followed by a plateau which finally yields a good estimation $\alpha$ small enough but larger than the typical first-neighbour distance. As an example, in the inset of Fig. SI.2a we show the behaviour of the total covered area, estimated via Delaunay triangulation[1], for a set of Poisson-distributed points in a square domain of side $L = 20$ where we removed two semicircles of radius $r_1 = 3$ and $r_2 = 8$. As clear from the plot, below a value of $\alpha \approx r_1$ the covered area converges to the analytical value, $L^2 - \pi^2(r_1^2 + r_2^2)/4$, provided the number of points is large enough.

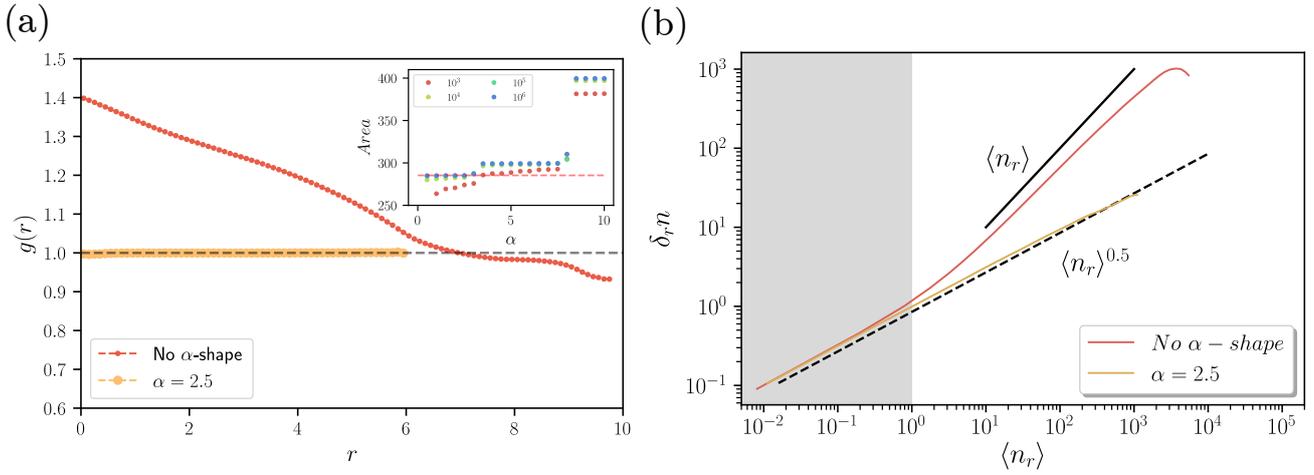

**Figure SI.2.** Effects of borders on density correlation and fluctuations. For a set of $N = 10^5$ points Poisson-distributed in a $20 \times 20$ square with two empty semicircles of radii $r_1 = 3$ and $r_2 = 8$ we compute (a) the pair correlation function $g(r)$ and (b) Taylor's law, using the borders as defined by the CH (red) or $\alpha-$shapes for $\alpha = 2.5$ (orange). The black dashed line is the theoretical results, $g(r) = 1$ and $\delta_r n_r = \langle n_r \rangle^{1/2}$. Inset of (a): covered area versus $\alpha$ for different number of points included in the same set as main panel. Notice that the covered area $A_\alpha$ converges to the analytical value (dashed line) for $\alpha < 3$ when the number of points $N$ is large enough (in this case $N \gtrsim 10^4$).

The main panel of Fig. SI.2a shows the effects of using the different boundaries on the computation of the pair correlation function (PCF), $g(r)$. When using the convex hull, $g(r)$ is found to be greater than 1 at small scales ($r < 6$) and converge to $\approx 1$ at large scales. This is a spurious "clumping" effect due to an incorrect choice of borders. In contrast, when using $\alpha$-shape, the expected result $g(r) = 1$ is found.

In Fig. SI.2b we show how biases induced by the incorrect identification of the borders impact on the evaluation of the density fluctuations, i.e. Taylor's law (TL). Without $\alpha$-shapes, i.e. using the borders of the square, one would have found strong deviations of TL exponent from the Poisson value $\gamma = 1/2$. Indeed, when failing to exclude the empty area we observe a power law with exponent $\gamma \approx 1$, which via Eq. (3.3) of main text would imply fluctuations that do not decay with the scale of observation. Instead, with the proper borders (i.e. removing the empty semicircles from the computation), and using the Hanisch method for avoiding border bias, we obtain the expected exponent $\gamma = 1/2$. The main reason for such spurious results is that empty cells are those which typically increase the fluctuations, so one should ensure, at each scale $r$, to include only the empty or semi-empty cells which really belong to the set of points. We conclude observing that at very small scales (gray shaded area in Fig. SI.2b) $\gamma = 1/2$ also with the wrong borders provided $\langle n_r \rangle < 1$, this is the trival regime discussed in Sec. 3 of the main text.

### SI-1.1: Area and density estimation in BCI using the $\alpha$-shapes

The dependence of covered area or density with $\alpha$ is not always as clear as in the inset of Fig. SI.2a. In actual BCI data many scales can be involved, and the step-like behaviour of the covered area is more an exception than a rule. This is particularly

---
[†]https://it.mathworks.com/help/matlab/ref/alphashape.html

**2/10**

clear from Figs. SI.3a-b where we show, for *H. prunifolius* and *T. panamensis* the dependence on $\alpha$ of the fraction of empty area $(A - A_\alpha)/A$ (where $A$ and is the area of the rectangular BCI plot and $A_\alpha$ that determined with the $\alpha$-shape algorithm). We observe an almost continuous and monotonic decay with $\alpha$ until the CH limit, in which no concavities are identified. Figures SI.3c-d show that the same applies to the relative difference between the $\alpha$-shape estimated density $\rho_\alpha = N/A_\alpha$ and the naive estimation, $\rho_0 = N/A$.

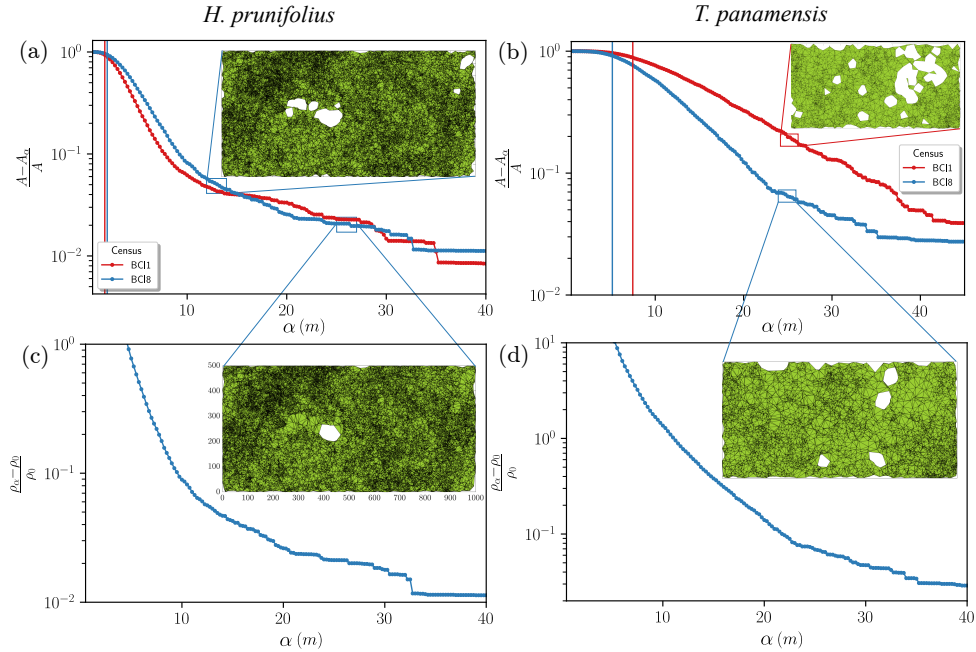

**Figure SI.3.** Estimation of covered area and density with $\alpha$-shapes. (a,b) Fraction of empty area $(A - A_\alpha)/A$ vs. $\alpha$, where $A$ is the area of the entire BCI plot and $A_\alpha$ the area with the concavities eliminated via $\alpha$-shape. (c,d) Relative difference between the naive density estimation $\rho_0$ and the density $\rho_\alpha$ computed using the area obtained with $\alpha$-shape vs. $\alpha$. Panels (a,c) correspond to *H. prunifolius* and (b,d) to *T. panamensis*. Note that no clear plateau can be identified (compare with the inset of Fig. SI.2a). For the fraction empty area we show results for the first census (red points) and the last one (blue points). Vertical bars indicate the interparticle distance for each census. Insets (a, c): area covered by *H. prunifolius* in census 8 for two values of $\alpha$ larger than the interparticle distance. Insets (b,d): area covered by *T. panamensis* for the same value of $\alpha$ in census 1 (b) and census 8 (d). The covered area depicted in the inset and its numerical value were obtained with Delaunay triangulation.

In the absence of clear plateaus, one is forced to use a more subjective criterion to fix the value of $\alpha$. Consider for instance the difference in covered area for two different values of $\alpha$ for *H. prunifolius* in the insets of Figs. SI.3a,c. To choose $\alpha$ we proceeded as follows: first, we always required $\alpha$ to be significantly larger than the mean distance between nearest-neighbour trees, to avoid fragmentation of the system (otherwise the covered area would be unreasonably small, see Figs. SI.3a-b). Inter-particle distances are plotted with vertical bars in the main panels of Figs.SI.3a-b. Second, we have taken $\alpha$ small enough to assure the identification of empty areas in the spatial distribution of different species, which were always checked by visual inspection. Further, to be more systematic, we also selected $\alpha$-values when the continuous decay of the estimated area changes its curvature.

As seen in the insets of Fig. SI.3, for *H. prunifolius* (see also Fig. 2 of main text) and *T. panamensis* such criterion provides a reasonable estimation of the covered area. In particular, for *H. prunifolius* our choice identified the empty region depicted in Fig. 2 of the main text (and in the inset of Fig. SI.3a) which actually corresponds to swampland region unfavorable to the establishement of this species[2]. Unfortunately, for other species we do not have additional information of this kind to check the validity of the choice of $\alpha$. Comparing the insets of Figs. SI.3b,d corresponding to the area covered by *T. panamensis* for the same value of $\alpha$ but for the first and last census, one can appreciate how some of the holes present in census 1 were filled in census 8, meaning that this species experienced an expansion in between the two censuses, whose consequences for the PCF are discussed in the main text (in particular Fig. 4d).

In Fig. SI.4 we show the border trees of the tree patterns for the ten most abundant species in BCI (which account for about half of all the trees present) together with the value of $\alpha$ we used. Finally, Fig. SI.5 shows the evolution of the borders in three different censuses for the contracting species *P. cordulatum* and *P. armata*. Notice that *P. cordulatum* seems to loose

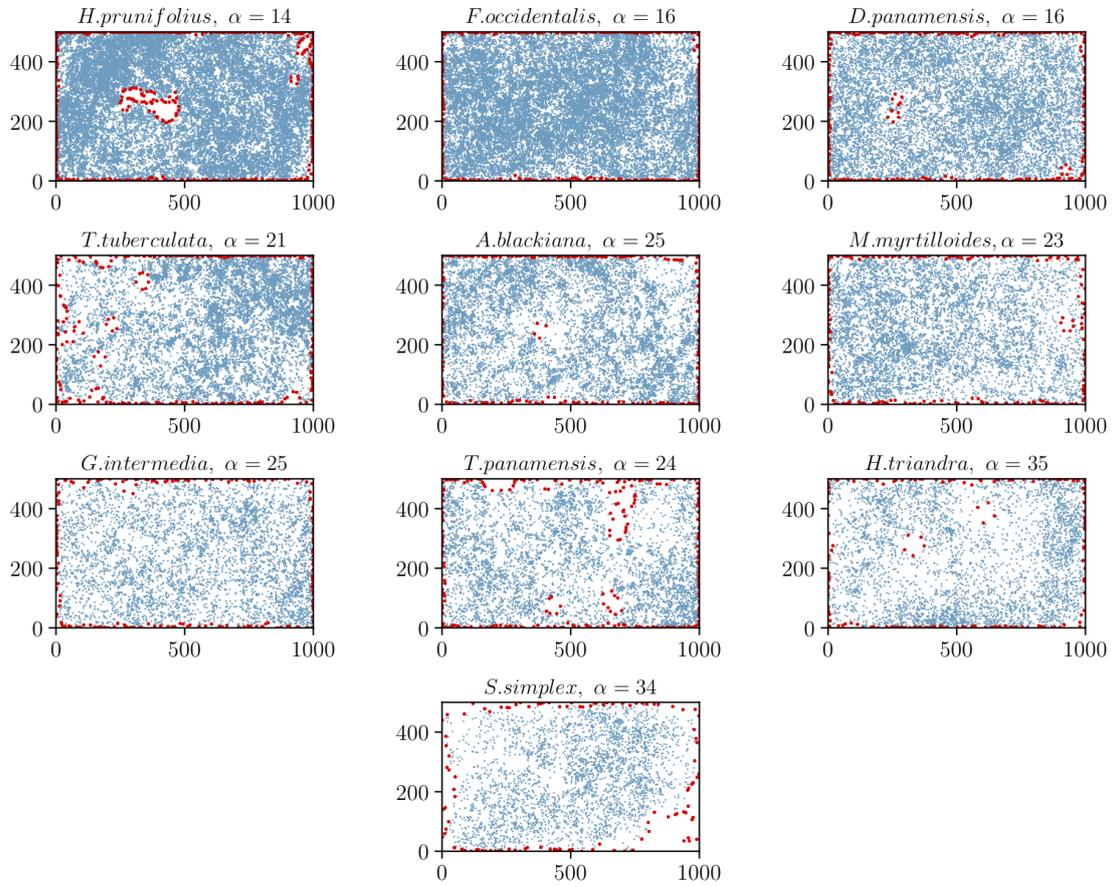

**Figure SI.4.** Border trees (red) identified with $\alpha$-shapes for the given values of $\alpha$ for the ten most abundant species in BCI.

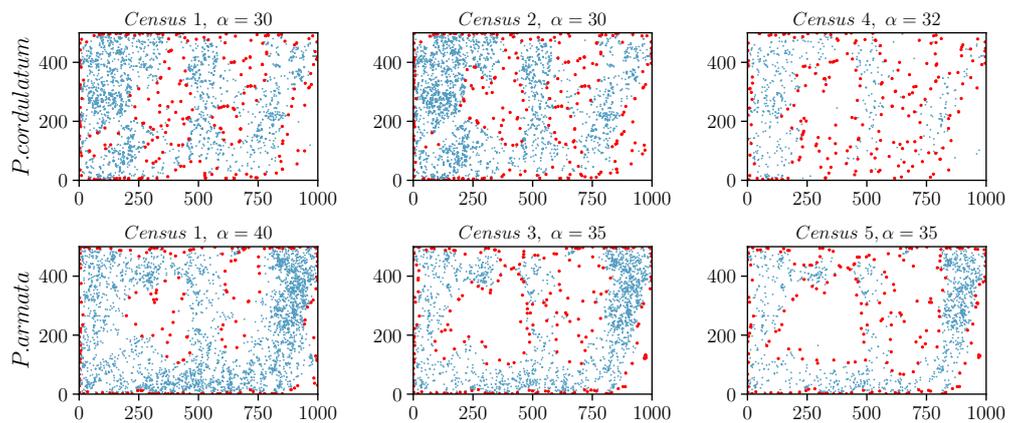

**Figure SI.5.** Border trees (red) identified with $\alpha$-shapes for the given values of $\alpha$ for two contracting species *P. cordulatum* (upper panels) and *P. armata* (lower panels) in different censuses.

structure (and as time progress looks more and more homogeneous), unlike *P. armata*. This is reflected in the behavior of density correlations (Fig. 5 of main text) and fluctuations (Fig.SI.8 and SI.9).

## SI-2: Density correlations and fluctuations at the single species level

As discussed in the main text, the most abundant species in Barro Colorado, while coherently showing evidence of clumping at small scales ($g(r) > 1$), clearly display very different behaviours at large scales. In the main text we mainly discussed three representative species. Here, we analyze those of Fig. SI.4, examining the effects of using $\alpha$-shapes on both the PCF (Fig. SI.6) and Taylor's law (Fig. SI.7).

Figure SI.6a shows the PCF computed using the rectangular plot borders. For all species we observe clumping at short distances, but at larger scales behaviours range from clear anticorrelation ($g(r) < 1$) at intermediate distances (see e.g. *H. triandra*), to species displaying a large scale plateaux at values either larger than 1 (e.g. *F. occidentalis* and *S. simplex*) or very near 1 (as e.g. *H. prunifolius* and *G. intermedia*). When using $\alpha$-shapes to find the border, as shown in Fig. SI.6b, some of the species recover a plateaux to 1 *(T. panamensis)* and anticorrelations essentially disappear, but still a quite large variability in the large scale behavior persists for other species.

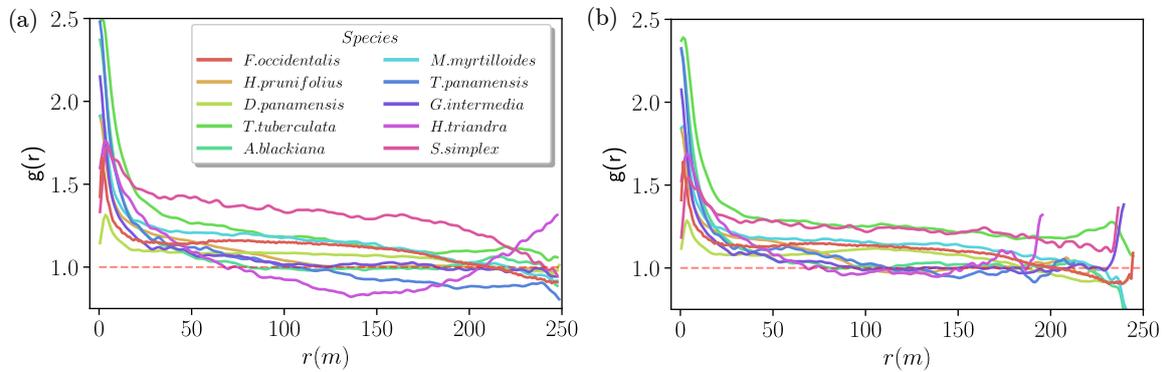

**Figure SI.6.** Pair correlation function $g(r)$ computed without (a) and with (b) $\alpha$-shapes for the ten most abundant species in BCI (distinguished by colour). The values of $\alpha$ used for each species and the resulting borders can be read in Fig. SI.4. Red dashed line shows the theoretical expectation $g(r) = 1$ for a completely random distribution of points.

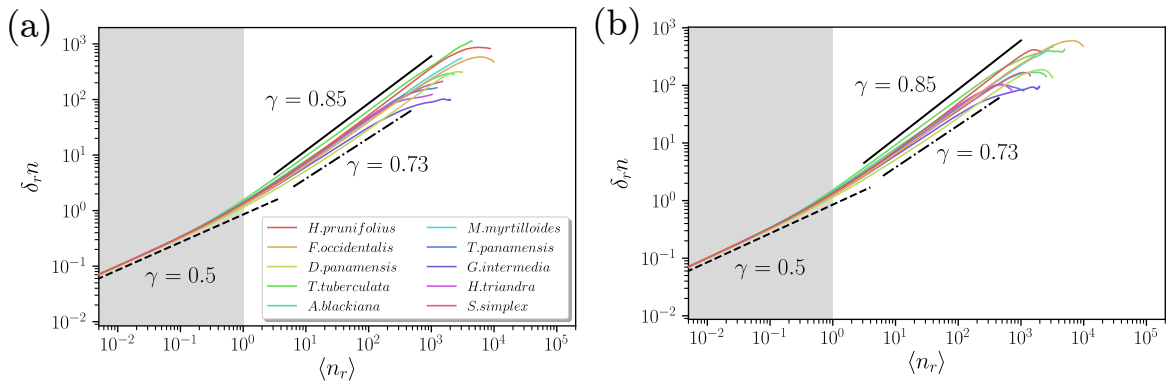

**Figure SI.7.** Density fluctuations measured via Taylor's law, i.e. looking at how the root mean square deviations of number of trees in cells of size $r$, $\delta_r n$, changes with mean $\langle n_r \rangle$, without (a) and with (b) $\alpha$-shapes. Dashed lines are the theoretical expectation, $\gamma = 1/2$, for a completely random process, while the solid and dashed-dotted ones bracket the (fitted) limiting value of the exponent $\gamma$. The grey shaded area indicates the region of scales below the mean distance between neighbouring trees, where $\langle n_r \rangle < 1$.

The fact that anti-correlations disappear means that some voids have been removed (this is likely the case of *H. triandra*, see e.g. Fig. SI.4). As briefly discussed in the main text, this points out a delicate issue when using methods such as $\alpha$-shapes to find borders: anticorrelations are not always spurious features of a tree distribution, so that when removing them one should be aware that one might be masking a genuine process responsible for them. On the other hand, in cases such as *T. panamensis*, which displays mild anticorrelation at large distances in Fig. SI.6a (see also main text Fig. 4d), the most likely explanation is that the species is experiencing an expansion (see also the insets of Fig. SI.3b-d).

Moving on to density fluctuations via Taylor's law (Sec. 3 of main text), we show in Fig. SI.7 how the root mean square deviations of number of trees in cells of size $r$, $\delta_r n$, changes with mean $\langle n_r \rangle$. We observed that apart from small inessential quantitative changes, the TL is not sensitive to using $\alpha$-shapes or not. This is somehow surprising for *H. prunifolius,f* for which a large hole in the domain is present. As for the quantitative aspects, we found that besides the trivial convergence to $\gamma = 1/2$ below the interparticle distance (gray shaded area), at large scales the behavior is always anomalous ($\gamma > 1/2$) for all species, with $\gamma$ varying in the range $[0.73 : 0.85]$. The effect of using $\alpha$-shapes or not is to produce small changes in the exponent of the single species, but the range of values remains basically the same.

In Figs. SI.8 and SI.9, we show density fluctuations for the two contracting species *P. cordulatum* and *P. armata*, showing the effect of using $\alpha$-shapes. We observe different behavior for the two species. For *P. cordulatum* we find strong dependence on the definition of the border, especially in census 4. In particular, $\gamma$ seems to approach the value $1/2$ indicative of a tendency of recovery of homogeneity. This seems to be confirmed by the behavior of the PCF shown in Fig.5a of the main text and with the visual impression from the top panels of Fig. SI.5, which show a qualitative change of the tree distribution in census 4 with respect to the previous two. For *P. armata* we found that $\gamma$ changes from 0.9 to 0.85, but there is no tendency to approach a homogeneous process.

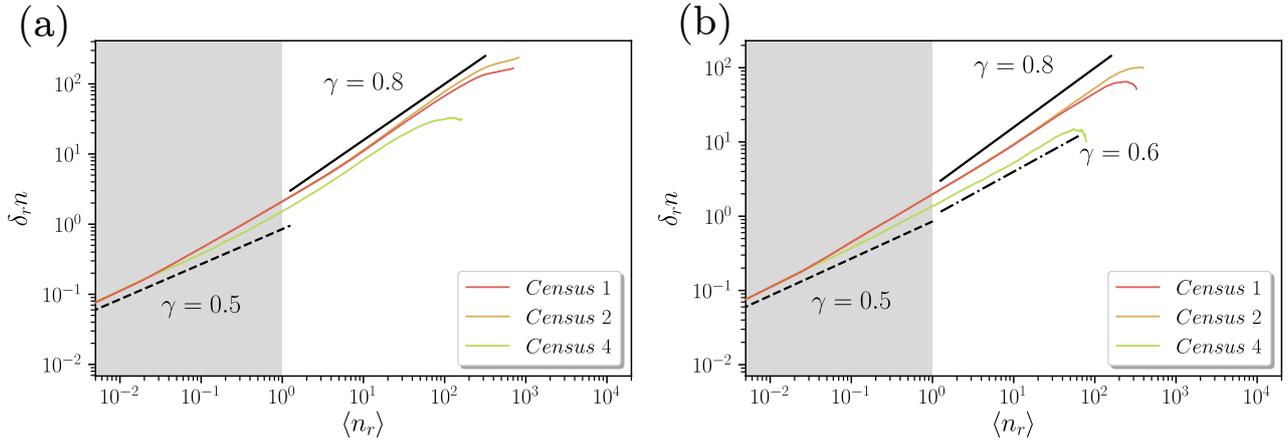

**Figure SI.8.** Taylor's law computed without (a) and with (b) $\alpha$-shapes for *P. cordulatum* in censuses 1 (with $\alpha = 30$), 2 ($\alpha = 30$) and 4 ($\alpha = 32$). Lines indicate the slopes as labeled. The grey shaded area indicates the region of scales below the mean distance between neighbouring trees, where $\langle n_r \rangle < 1$.

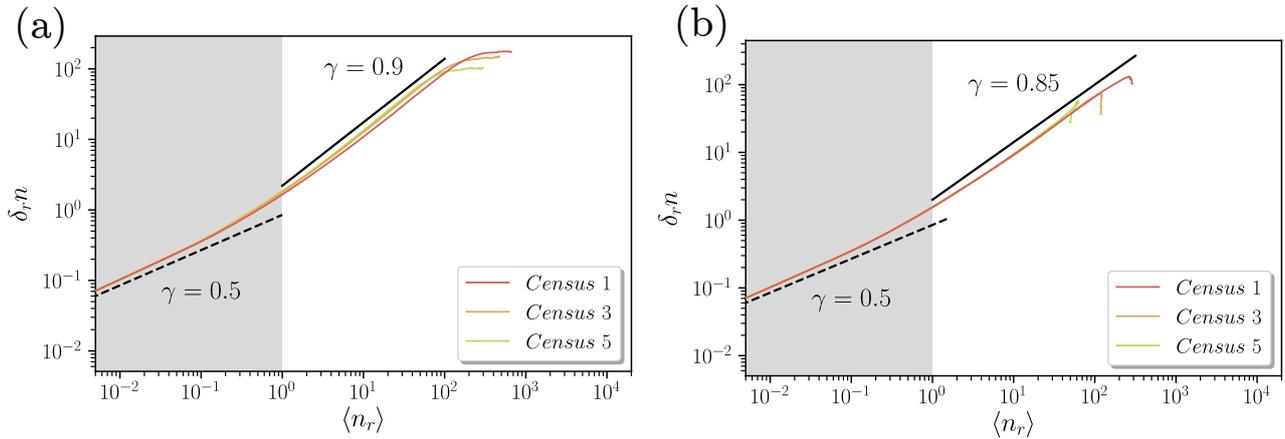

**Figure SI.9.** As Fig. SI.8 but for *P. armata* in censuses 1 ($\alpha = 40$), 3 ($\alpha = 35$) and 5 ($\alpha = 35$).

## SI-3: Dual representation of the spatially explicit neutral model

The multispecies voter model has been simulated exploiting its duality with a system of coalescing random walkers with a killing rate[3–5]. The main advantage of this method is that it allows for very fast simulations to produce independent realization of sample patterns at the (non-equilibrium) stationary state virtually free from boundary effects such as that can be introduced by periodic boundary conditions (typically employed when simulating in the forward representation described in the main text).

The dual process is built as follows. At the beginning, each lattice site is filled with a random walker. The dual process proceeds backwards in time to reconstruct the ancestry of the species. At each discrete (backward) time step, with probability $1 - \nu$, a randomly chosen walker is moved to a different site (which can be outside the sampled domain, since the lattice is infinite though we only observe a finite portion), chosen according to a distribution which depends on the distance from the original site $r$ —i.e. the dispersal kernel $P(\mathbf{r})$, which we have chosen to be Gaussian centred in the original site and with standard deviation $\sigma$. If the landing site is occupied, the two walkers coalesce and one of them is removed, keeping trace of the coalescing partner. With complementary probability $\nu$, the randomly chosen walker is killed. This corresponds to a speciation event in the forward in time description. The simulation proceeds until only one walker is present. Finally, having stored the whole tree of coalescences and knowing which walker was killed one can trace back the entire genealogical tree of a species up to the speciation event that originated it. Then, having labelled each walker in such a way to be able to identify its initial position, one can assign to each site of the lattice. Since the number of walkers decreases at each coalescence or killing event, the simulation time becomes faster as time proceeds. Of course, this procedure can be used only if one is interested in the static, long-term, properties of the model.

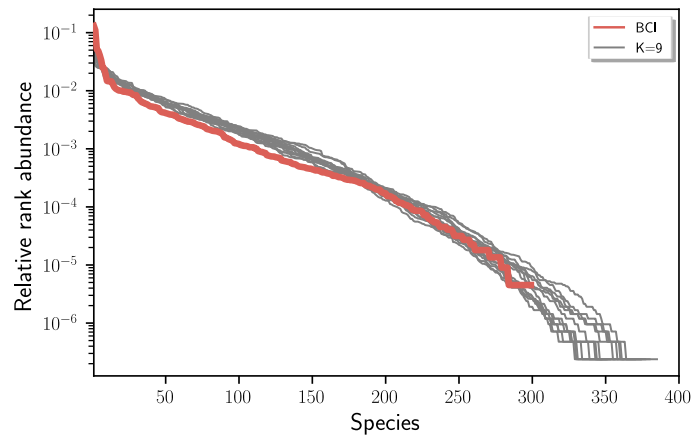

**Figure SI.10.** Rank abundance curves (grey lines) for different realizations of the MVM with $\sigma = 9$, $\nu = 3.8 \cdot 10^{-6}$ and the rank abundance of the 8th census of Barro Colorado (red line).

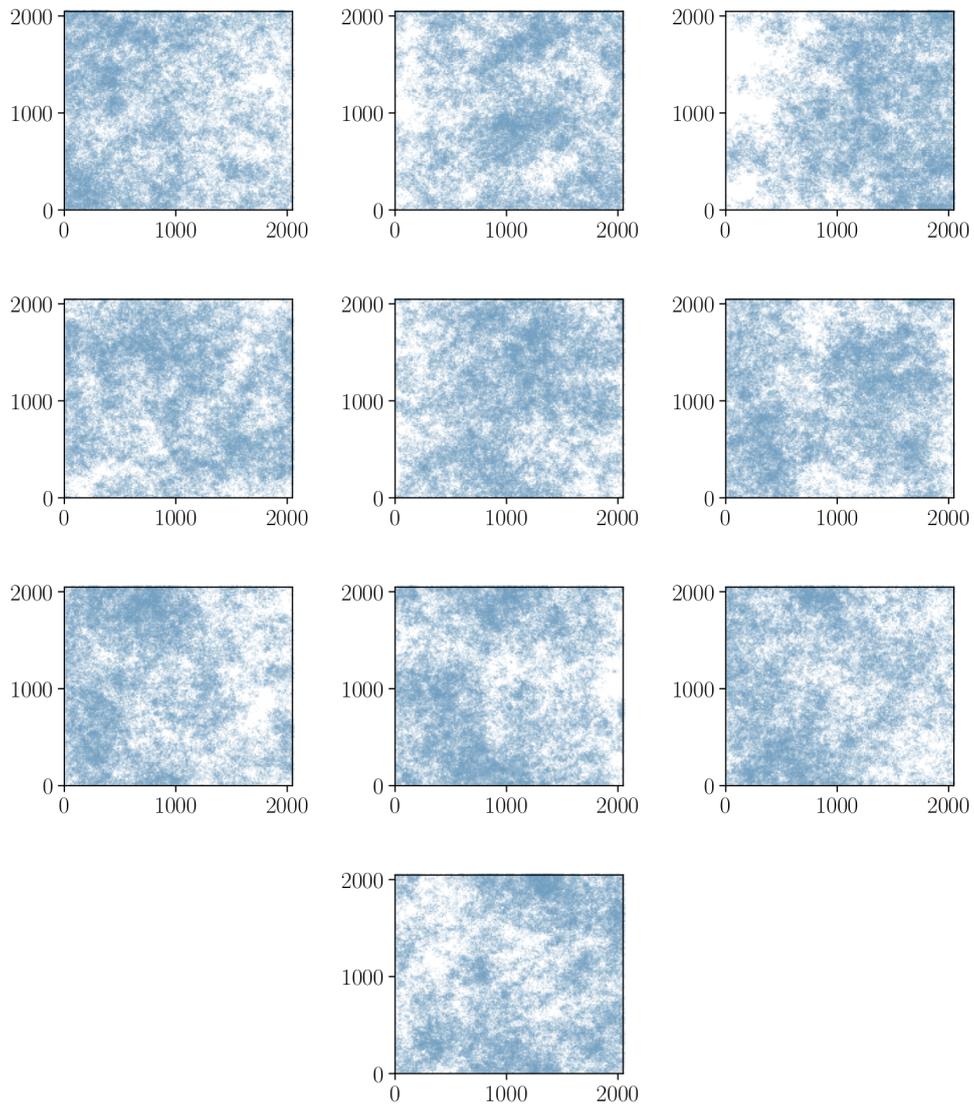

**Figure SI.11.** Spatial patterns for different species in the MVM. Parameters: $\sigma = 9$, $\nu = 3.8 \cdot 10^{-6}$ and $N_i \in (142000 - 157000)$.

**SI-3.1: Effects of $\alpha-shapes$ method**

In the case of the above described spatially explicit neutral model we have a complete epistemic knowledge of the system and, in particular, we know that the (eco-)system is homogeneous in the sense that all points occupied by trees are equivalent a priori. This means that there are no reasons to consider borders besides the square where the process is simulated, which should be regarded as a sample of an infinite system. However, as clear from Fig. SI.11 voids are present due to the competition between species. Thus, for the sake of completeness and comparison with BCI, it is worth considering the effect of $\alpha$-shapes also within the patterns generated with the neutral model.

In Fig. SI.12 we show the border of the spatial patterns for some selected species in a neutral environment together with the value of $\alpha$ we used.

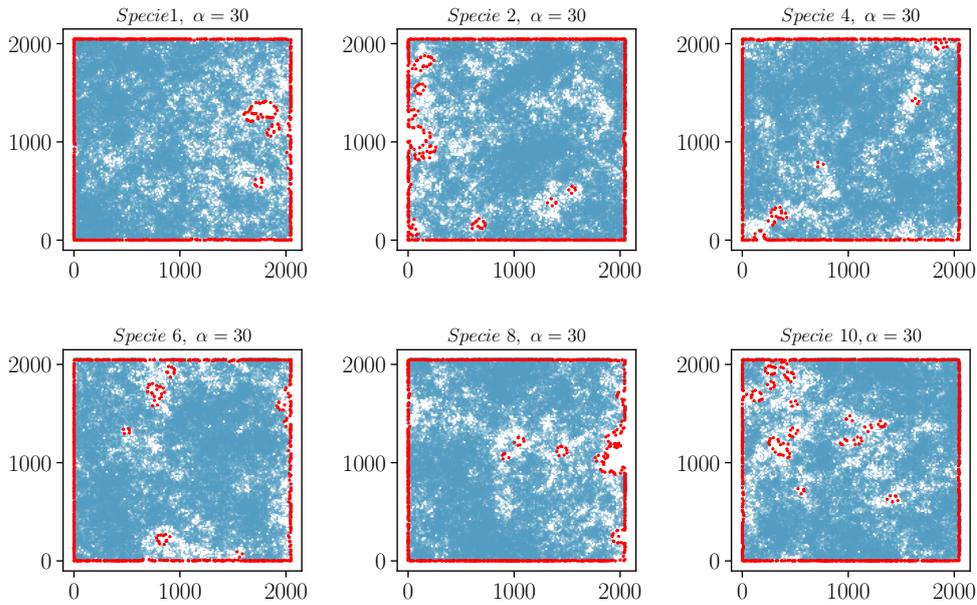

**Figure SI.12.** Spatial distribution for some selected simulated species (with $N_i \approx 1.4$–$1.6 \cdot 10^5$ trees) with the neutral model in a lattice of side $L = 2048$. Borders are outlined in red. The internal borders can be clearly appreciated when $\alpha-shapes$ are employed.

Regarding to the g(r), some differences between both cases (see Fig.SI.13a and Fig.SI.13b) can be observed. Firstly, the behavior at short scales is not affected by the $\alpha$-shape method. That is, clumping behavior is completely robust against changes of the borders. For larger scales, the effect of $\alpha-shapes$ consists of mainly removing anticorrelations. However, we stress again, in this case there is no plausible reason –as also discussed above and in the main text– to remove such anticorrelations in the pattern generated by this model.

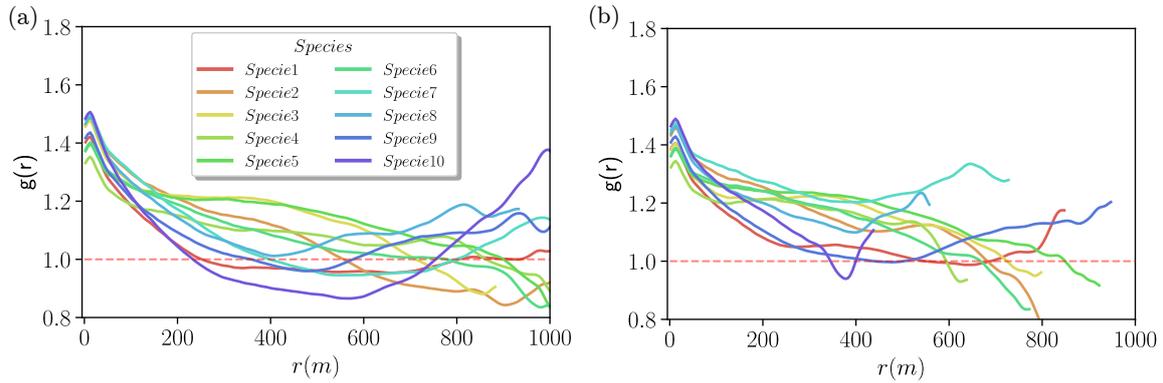

**Figure SI.13.** Pair correlation function for ten selected simulated species with the neutral model (a) without $\alpha-shapes$ and (b) employing the $\alpha-shape$ method with a predefined radius $\alpha = 30$. For some of the species anticorrelations at mid-scales clearly disappears (e.g. Specie10).

For the sake of completeness, we also investigated the possible role of boundaries in the behavior of spatial density fluctuations, i.e. in the behavior of the Taylor's law. As shown in Fig.SI.14a and Fig.SI.14b, we have not found any bias on the value of the exponent $\gamma$ when the $\alpha-shape$ method is employed. Thus, we can safely say that –for reasonable values of *alpha*, larger than the interparticle distance– large spatial fluctuations do not depend on the selection of the borders for the neutral model, at least for the most abundant species there represented.

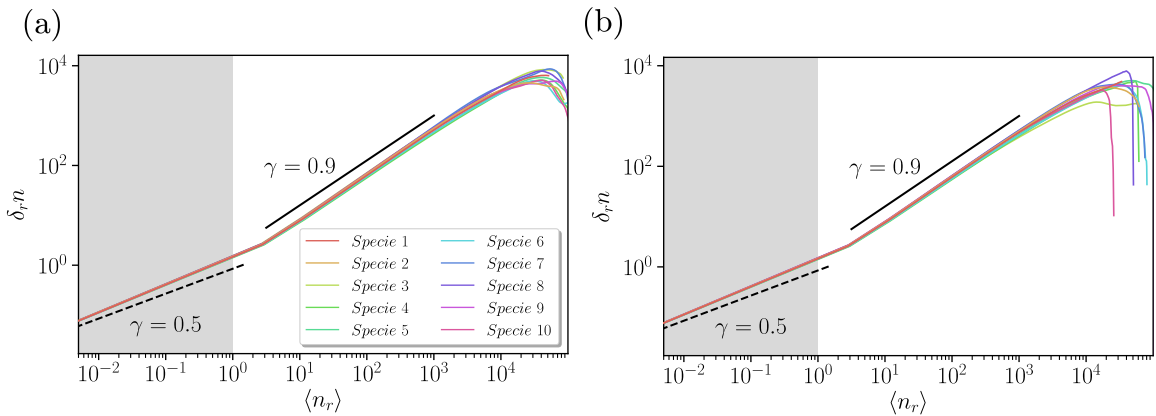

**Figure SI.14.** Spatial density fluctuations for the ten selected simulated species with the neutral model (a) without $\alpha-shapes$ and (b) employing the $\alpha-shape$ method with a predefined radius $\alpha = 30$. There is no qualitatively change between both cases.



\end{document}